\documentclass[nonacm,format=manuscript,authorversion,screen]{acmart}

\usepackage{xparse}
\usepackage{xpatch}
\usepackage{etoolbox}
\usepackage{environ}
\usepackage{lineno}
\usepackage{multicol}
\usepackage{xspace}

\renewcommand\footnotetextcopyrightpermission[1]{}
\setcopyright{none}

\usepackage[utf8]{inputenc}

\colorlet{keyword}{blue!25!black}
\colorlet{literal}{red!40!black}
\colorlet{type}{green!40!black}
\colorlet{comment}{green!20!black}

\usepackage[british]{babel}

\usepackage{stmaryrd}
\usepackage{mathtools}
\usepackage{esvect}
\renewcommand{\vec}{\vv}
\allowdisplaybreaks

\usepackage{placeins}

\usepackage[inline]{enumitem}

\usepackage{tikz}
\usetikzlibrary{calc,arrows,positioning,shapes.multipart,arrows.meta,decorations.pathmorphing}
\tikzset{
  squiggly/.style = {
    line join=round,
    decorate, decoration={
      zigzag,
      segment length=4,
      amplitude=.9,post=lineto,
      post length=2pt}
  }
}

\usepackage{hyperref}

\usepackage[capitalise,nameinlink,noabbrev]{cleveref}

\makeatletter
\newcommand{\customlabel}[4][0]{%
	\protected@write\@auxout{}{\string\newlabel{#3}{{#4}{\thepage}{#4}{#3}{}}}%
  \protected@write\@auxout{}{\string\newlabel{#3@cref}{{[#2][#1][#1]#4}{\thepage}{}{}{}}}%
}
\NewDocumentCommand\declarecrefname{s m m m m m}{
	\Crefname{#2}{#5}{#6}
	\IfBooleanTF{#1}{
		\crefname{#2}{#3}{#4}	
	}{
		\if@cref@capitalise
			\crefname{#2}{#5}{#6}	
		\else
			\crefname{#2}{#3}{#4}	
		\fi
	}
}
\makeatother

\hypersetup{
	hidelinks,
	final,
}

\makeatletter
\newcommand\autoflow@gobblepars{%
	\@ifnextchar\par%
		{\expandafter\autoflow@gobblepars\@gobble}%
		{}}

\newcommand\autoflow@AND{\unskip%
	\hskip1ptplus2pt%
	\cleaders\copy\autoflow@ANDbox\hskip\wd\autoflow@ANDbox%
	\hskip1ptplus2pt%
	\ignorespaces\autoflow@gobblepars}

\newsavebox\autoflow@ANDbox

\newcommand{\setautoflowsep}[1]{\sbox\autoflow@ANDbox{#1}}
\setautoflowsep{\qquad}

\newenvironment{autoflow}{%
	\begingroup%
	\def\AND{\autoflow@AND}
	\parskip=0pt%
	\parindent=0pt%
	\par%
	\ignorespaces%
	}{\endgroup\ignorespacesafterend}

\makeatother

 \usepackage{proof-dashed}

\NewDocumentCommand{\rname}{o m}{%
	\ensuremath{\left\lfloor\mspace{-2mu}%
	\IfNoValueF{#1}{\textnormal{\scshape #1}\middle|}%
	\textnormal{\scshape #2}%
	\mspace{-2mu}\right\rceil}}
\NewDocumentCommand{\rlabel}{m m}{{%
	\customlabel{infrule}{#2}{#1}%
	\hypertarget{#2}{#1}}
  }
\makeatletter
\NewDocumentCommand{\infrule}{o o m m}{%
	\begingroup%
	\let\and\quad%
	\renewcommand{\arraystretch}{\infrulestretch}%
	\setlength\arraycolsep{1pt}%
	\def\infrule@P{#4}%
	\def\infrule@C{#3}%
	\settoheight{\infrule@Ph}{\mbox{\math\displaystyle\begin{array}{c}\smash{A}\infrule@P\end{array}\endmath}}%
	\settoheight{\infrule@Ch}{\mbox{\math\displaystyle\begin{array}{c}\infrule@C\end{array}\endmath}}%
	\raisebox{\dimexpr \infrule@Ph - \infrule@Ch\relax}{$\displaystyle%
		\begin{array}{c}%
			\infrule@P%
			\\\hline%
			\infrule@C%
		\end{array}%
	$}~\IfNoValueF{#1}{\hspace{-2pt}\raisebox{.1ex}{\scalebox{.9}%
		{\IfNoValueTF{#2}{#1}{\rlabel{#1}{#2}}}}}%
	\endgroup%
	\AND%
}
\newlength{\infrule@Ph}
\newlength{\infrule@Ch}
\newenvironment{infrules}{%
	\center%
	\begingroup%
		\setautoflowsep{\infrulehsep}%
		\baselineskip=\infrulelineskip%
		\autoflow%
	}{\endautoflow\endgroup\endcenter}
\makeatother
\newcommand*{\infrulestretch}{1.4}
\newcommand*{\infrulehsep}{\qquad}
\newcommand*{\infrulelineskip}{\bigskipamount}

\declarecrefname{infrule}{rule}{rules}{Rule}{Rules}

\usepackage[final]{listings}

\lstdefinelanguage{Choral}{
  morekeywords={class,else,extends,if,implements,interface,new,null,private,%
      protected,public,abstract,final,static,return,super,this,try,catch,void, %
      enum,switch,case,default,throw,throws,with,forall,in},%
  otherkeywords={>>,::,@},
  sensitive,%
  escapechar=¦,
	morecomment=[l]//,%
	morecomment=[s]{/*}{*/},%
	morestring=[b]",%
  morestring=[b]',%
  moredelim=[is][]{\#}{\#}%
}[keywords,comments,strings]%

\lstdefinelanguage{comperr}{
  morekeywords={class,else,extends,if,implements,interface,new,null,private,%
      protected,public,abstract,final,static,return,super,this,try,catch,void, %
      enum,switch,case,default,throw,throws},%
  otherkeywords={>>,@},
  sensitive,%
	morecomment=[l]//,%
	morecomment=[s]{/*}{*/},%
	morestring=[b]",%
  morestring=[b]',%
  moredelim=[is][]{\#}{\#},
  moredelim=[is][\bfseries\color{sred}]{~}{~},
  moredelim=[is][basicstyle]{/~}{~/}
}[keywords,comments,strings]%

\usepackage[scaled=.85]{FiraMono}

\definecolor{sbase03}{HTML}{002B36}
\definecolor{sbase02}{HTML}{073642}
\definecolor{sbase01}{HTML}{586E75}
\definecolor{sbase00}{HTML}{657B83}
\definecolor{sbase0}{HTML}{839496}
\definecolor{sbase1}{HTML}{93A1A1}
\definecolor{sbase2}{HTML}{EEE8D5}
\definecolor{sbase3}{HTML}{FDF6E3}
\definecolor{syellow}{HTML}{B58900}
\definecolor{sorange}{HTML}{CB4B16}
\definecolor{sred}{HTML}{DC322F}
\definecolor{smagenta}{HTML}{D33682}
\definecolor{sviolet}{HTML}{6C71C4}
\definecolor{sblue}{HTML}{268BD2}
\definecolor{scyan}{HTML}{2AA198}
\definecolor{sgreen}{HTML}{859900}

\lstdefinestyle{solarized-light}{
  frame=none,
  breaklines=true,
  showstringspaces=false,
  tabsize=1,
  columns=fixed,
  mathescape=true,
  extendedchars=true,
  backgroundcolor=\color{sbase3},
  keywordstyle=\bfseries\color{sbase01},
  keywordstyle=[2]\bfseries\color{sbase01},
  stringstyle=\color{sviolet},
  numberstyle=\color{sviolet},
  identifierstyle=\color{sbase03},
  commentstyle=\color{sgreen},
  basicstyle=\color{sbase03}\ttfamily\lst@ifdisplaystyle\footnotesize\fi,
  moredelim=[is][\color{sblue}]{\#}{\#},
}

\usepackage[most]{tcolorbox}
\tcbuselibrary{listings}

\NewDocumentCommand{\newlang}{o m m m}{%
  \IfNoValueTF{#1}{\def\n{#2}}{\def\n{#1}}
  \expandafter\NewDocumentCommand\csname \n listing\endcsname{s O{} O{}}{%
	  \def\WithoutTitle{\tcblisting{
	          enhanced, %

	          before skip=\abovedisplayskip,
	          after  skip=\belowdisplayskip,
	          sharp corners=all,
	          boxrule=2pt,
	          boxsep=-.7em,
	          colframe=#4,
	          colback=sbase3,
	          title={},
	          listing only,
	          listing options={language=#2,numbers=left,style=solarized-light,##3},
	          ##2
 	  	}}%
  	\def\withTitle{\tcblisting{
         enhanced, %
         before skip=\abovedisplayskip,
         after  skip=\belowdisplayskip,
         sharp corners=all,
         boxrule=2pt,
         boxsep=-.7em,
         colframe=#4,
         colback=sbase3,
         detach title,
         finish={\node[anchor=south east, font=\footnotesize\itshape,
         text=sbase3,fill=#4] at (frame.south east) {#3 Code};},
         title={},
         listing only,
         listing options={language=#2,numbers=left,style=solarized-light,##3},
         ##2
   }}%
   \IfBooleanTF{##1}{\WithoutTitle}{\withTitle}}
  \expandafter\def\csname end\n listing\endcsname{\endtcblisting\noindent}
  \expandafter\def\csname\n\endcsname{\lstinline[language=#2,style=solarized-light]}
}

\newlang[choral]{choral}{Choral}{sblue}
\newlang[comm]{java}{Generated}{syellow}
\newlang[local]{java}{Local}{sbase01}
\newlang[java]{java}{Java}{sbase1}

\newenvironment{snippetbox}[1][]{%
	\begin{tcolorbox}[
		colback=black!5,colframe=black!30,
		boxsep=-1pt,
		boxrule=.5pt,
		sharp corners,
    breakable,
		#1
	]}{%
		\end{tcolorbox}\noindent\ignorespacesafterend%
	}
\newenvironment{snippet}[1][]{%
	\begin{snippetbox}%
  \edef\cLN{\thelinenumber}%
	\begingroup%
	\renewcommand*{\\}[1][]{\endmath\par\vskip2pt\math\displaystyle}%
	\renewcommand*{\indent}{\quad}%
	\providecommand*{\commentline}[1]{\hspace{5pt}\text{\small\color{comment}/\kern-2pt/\ ##1}}%
	#1%
	\begin{internallinenumbers}%
	\resetlinenumber[1]%
	\setlength\linenumbersep{1pt}%
  \math\displaystyle%
}{%
	\endmath%
	\end{internallinenumbers}%
	\endgroup%
  \addtocounter{linenumber}{\cLN}%
	\end{snippetbox}%
	\noindent\ignorespacesafterend%
}

\newenvironment{snippet*}[1][]{%
	\begin{snippetbox}%
	\begingroup%
	\renewcommand*{\\}[1][]{\endmath\par\math\displaystyle}%
	\renewcommand*{\indent}{\quad}%
	\providecommand*{\commentline}[2][\qquad]{##1\text{//##2}}%
	#1%
	\math\displaystyle%
}{%
	\endmath%
	\endgroup%
	\end{snippetbox}%
	\noindent\ignorespacesafterend%
}

\newcommand*{\RC}{LC\xspace}

\newcommand{\defeq}{\triangleq}
\newcommand{\defiff}{\mathrel{\setbox0\hbox{$\iff$}%
    \rlap{\hbox to \wd0{\hss$\triangleq$\hss}}\box0}}

\newcommand*{\code}[1]{\text{\ttfamily\upshape #1}}
\newcommand*{\keyword}[1]{{\text{\color{keyword}\ttfamily\upshape #1}}}

\newcommand*{\literal}[1]{{\text{\color{literal}\ttfamily\upshape #1}}}

\newcommand*{\lbl}[1]{\textsc{#1}}

\newcommand*{\aspid}[1]{\IfBooleanF{#1}{\proc}}

\newcommand*{\cnil}{\mathbf{0}}

\newcommand*{\conc}{\keyword{;}\,}
\newcommand*{\ttto}
	{\mathrel{\keyword{\raisebox{.05ex}{-}\kern-0ex>}}}
\newcommand*{\ttfrom}
	{\mathrel{\keyword{<\kern-0ex\raisebox{.05ex}{-}}}}
\newcommand*{\tteq}
	{\mathrel{\keyword{:\raisebox{-.13ex}{=}}}}
\newcommand*{\ttTo}
	{\mathrel{\keyword{\raisebox{.0ex}{=}\kern-.0ex>}}}

\newcommand{\procdef}[3]{#1(#2) \keyword{\;=\;} #3}

\NewDocumentCommand{\epp}{m o}
	{\left\llbracket{#1}\right\rrbracket%
		\IfNoValueF{#2}{_{#2}}}

\let\merge\sqcup
\let\pruning\sqsupseteq

\newcommand{\proc}[1]{\mathsf{#1}}
\newcommand{\procs}[1]{\vec{\proc{#1}}}

\newcommand*{\apar}{\mathrel{\keyword{\bfseries |}}}

\newcommand{\newframepair}[5]{(#1,#2)^{\typef{#5}}\colon\proc{#3}\to \proc{#4}}
\newcommand{\gnewframepair}{\newframepair{k}{k'}{p}{q}{T}}
\newcommand{\newframe}[4]{\proc{#2}.(\proc{#3},#1)^{\typef{#4}}}
\newcommand{\gnewframe}{\newframe{k}{p}{q}{T}}
\newcommand{\newframel}[2]{(\proc{#1},#2)}

\newcommand{\send}[3]{\proc{#1}.#2!#3}
\newcommand{\gsend}{\send{p}{c}{s}}
\newcommand{\gssend}{\send{p}{(\proc{q},m)}{s}}
\newcommand{\sendl}[2]{#1!#2}
\newcommand{\gsendl}{\sendl{c}{s}}

\newcommand{\recv}[3]{\proc{#1}.#2?#3}
\newcommand{\grecv}{\recv{p}{c}{x}}
\newcommand{\gsrecv}{\recv{p}{(\proc{q},m)}{x}}
\newcommand{\recvl}[2]{#1?#2}
\newcommand{\grecvl}{\recvl{c}{x}}

\newcommand{\branch}[2]{#1\&\{#2\}}
\newcommand{\gbranch}{\branch{c}{l_1:B_i}_{i\in I}}

\newcommand{\init}[5]{(#1,#2)^{\typef{#5}}:\proc{#3}\to \proc{#4}}
\newcommand{\ginit}{\init{k}{k'}{p}{q}{T}}
\newcommand{\initl}[2]{#2:\proc{#1}}
\newcommand{\ginitl}{\initl{p}{k}}

\newcommand{\assign}[3]{\proc{#1}.#2:=#3}
\newcommand{\gassign}{\assign{p}{x}{s}}
\newcommand{\assignl}[2]{#1:=#2}
\newcommand{\gassignl}{\assignl{x}{s}}

\newcommand{\cond}[3]{\text{if }#1\text{ then }#2\text{ else }#3}
\newcommand{\gcond}{\cond{\proc{p}.b}{C_1}{C_2}}
\newcommand{\gcondl}{\cond{b}{B_1}{B_2}}
\newcommand{\gprocdef}{\procdef{X}{\procs p;\vec{k}}{C}}
\newcommand{\call}[3]{#1\tuple{#2;#3}}
\newcommand{\gcall}{\call{X}{\procs{p}}{\vec{c}}}
\newcommand{\cont}[2]{#1[#2]}
\newcommand{\gcont}{\procs{q}:\cont{X}{\procs{p};\vec{c}}.C'}

\newcommand{\cdefs}{\ensuremath{\mathcal{C}}}
\newcommand{\pdefs}{\ensuremath{\mathcal{B}}}
\newcommand{\tuple}[1]{\ensuremath{\langle #1 \rangle}}
\newcommand{\ltoc}[2]{\xrightarrow{#1}_{#2}}
\newcommand{\lto}[1]{\ltoc{#1}{\cdefs}}
\newcommand{\ltol}[1]{\ltoc{#1}{\pdefs}}
\newcommand{\disjoint}{\mathbin{\#}}
\newcommand{\eval}[4]{\ensuremath{{#2}(\proc{#3})\vdash{#1}\downarrow{#4}}}

\newcommand{\typef}[1]{\ensuremath{\mathsf{#1}}}
\newcommand{\sndt}[1]{{!}\typef{#1}}
\newcommand{\rcvt}[1]{{?}\typef{#1}}
\newcommand{\mrgf}{\Ydown}

\newcommand{\rbst}[5]{#2;#1\vdash\{#4\}\,{#3}\,\{#5\}}

\newcommand{\lbldefault}{\textsc{default}}
\newcommand{\fbuf}{\mathbf{fb}}
\newcommand{\fcnt}{\mathbf{fc}}

\newcommand{\con}{\fcnt}

\newcommand{\true}{\mathsf{true}}
\newcommand{\false}{\mathsf{false}}
\newcommand{\pn}{\mathsf{pn}}

\newcommand{\labels}{\mathcal{L}}

\title{A Promising Future: Omission Failures in Choreographic Programming}
\author{Eva Graversen}
\orcid{0000-0002-9430-4907}
\email{efgraversen@imada.sdu.dk}
\author{Fabrizio Montesi}
\orcid{0000-0003-4666-901X}
\email{fmontesi@imada.sdu.dk}
\author{Marco Peressotti}
\orcid{0000-0002-0243-0480}
\email{peressotti@imada.sdu.dk}
\affiliation{University of Southern Denmark}



\begin{document}
\theoremstyle{acmdefinition}
\newtheorem{remark}[theorem]{Remark}

\begin{abstract}
Choreographic programming promises a simple approach to the coding of concurrent and distributed systems: write the collective communication behaviour of a system of processes as a \emph{choreography}, and then the programs for these processes are automatically compiled by a provably-correct procedure known as endpoint projection.
While this promise prompted substantial research, a theory that can deal with realistic \emph{communication failures} in a distributed network remains elusive.

In this work, we provide the first theory of choreographic programming that addresses realistic communication failures taken from the literature of distributed systems: processes can send or receive fewer messages than they should (\emph{send} and \emph{receive omission}), and the network can fail at transporting messages (\emph{omission failure}).
Our theory supports the programming of strategies for failure recovery, and a novel static analysis (called \emph{robustness}) to check for delivery guarantees (at-most-once and exactly-once).

Our key technical innovation is a deconstruction of the usual communication primitive in choreographies to allow for independent implementations of the send and receive actions of a communication, while still retaining the static guarantee that these actions will correlate correctly (the essence of choreographic programming).
This has two main benefits.
First, each side of a communication can adopt its own failure recovery strategy, as in realistic protocols.
Second, initiating new communications does not require any (unrealistic) synchronisation over unreliable channels: senders and receivers agree by construction on how each message should be identified.
We validate our design via a series of examples -- including two-phase commit, which so far eluded choreographic programming -- and an implementation of our ideas in the choreographic programming language Choral.
\end{abstract}
\maketitle

\section{Introduction}
\label{sec:intro}
\paragraph*{Background: choreographic programming}
In the paradigm of choreographic programming~\cite{M13:phd}, programs express coordination plans for communicating processes. A choreographic programming language typically offers high-level communication primitives inspired by the Alice and Bob notation for security protocols~\cite{NS78}.
In particular, these languages extend the abstraction $\proc p \to \proc q$ -- read `process $\proc p$ communicates a message to process $\proc q$' -- with ways to express message payloads, computation, and control flow.

Given a choreography, a distributed implementation can be automatically compiled through a procedure known as endpoint projection (EPP) -- see \cref{fig:chor-prog}~\cite{CHY12,M23}.
EPP is key to the success of choreographic programming: it provides an escape from the challenge of separately writing compatible process programs, which is notoriously hard even for expert developers~\cite{LLLG16}.
This benefit, combined with the simplicity of the paradigm, has motivated a number of investigations into how choreographic programming languages can be understood, applied, and extended.
Recent examples include a full-fledged object-oriented choreographic programming language that extends Java~\cite{GMP24}, choreographic programming libraries for Haskell~\cite{SKK23} and Rust~\cite{KSZK23}, several mechanisations of choreographic programming theories~\cite{CMP23,HG22,PGSN22}, and the correct implementation of distributed cryptographic applications~\cite{AGRM24,ARGMS21}.

\pgfdeclarelayer{background}
\pgfdeclarelayer{foreground}
\pgfsetlayers{background,main,foreground}

\begin{figure}[t]
\centering
{
\begin{tikzpicture}[font=\footnotesize]
\node [
	rectangle, rounded corners, draw, outer sep=2pt,
	minimum width=2.7cm,
	label=above:Choreography,
	align=left,
] (chor) {$\proc a \mathbin{\texttt{->}} \proc b:x;$ \\ $\proc a \mathbin{\texttt{->}} \proc c:y;$ \\ $\proc b$ computes $z$; \\ $\proc b \mathbin{\texttt{->}} \proc c:z;$};

\node [
	rectangle, rounded corners, draw,
	dashed,
	right=0.5cm of chor,
	minimum width=2.7cm,
	align=center,
	fill=white
] (proj) {Endpoint \\ projection (EPP)};

\node [
	rectangle, rounded corners, draw, outer sep=2pt,
	right=1.5cm of proj,
	minimum width=2.7cm,
	label=above:Code for process $\proc b$,
	align=left,
] (B) {recv $x$ from $\proc a$;\\ compute $z$;\\ send $z$ to $\proc c$;};

\node [
	rectangle, rounded corners, draw, outer sep=2pt,
	above=.8cm of B,
	minimum width=2.7cm,
	label=above:Code for process $\proc a$,
	align=left,
] (A) {send $x$ to $\proc b$;\\ send $y$ to $\proc c$};

\node [
	rectangle, rounded corners, draw, outer sep=2pt,
	below=0.8cm of B,
	minimum width=2.7cm,
	label=above:Code for process $\proc c$,
	align=left,
] (N) {recv $y$ from $\proc a$; \\ recv $z$ from $\proc b$;};

\begin{pgfonlayer}{background}
\draw[-{Latex[length=1.5mm,width=3mm]},dashed] 
      (chor) -- (proj) -- ++ (2,0) |- (A);
\draw[-{Latex[length=1.5mm,width=3mm]},dashed] 
      (chor) -- (proj) -- ++ (2,0) |- (B);
\draw[-{Latex[length=1.5mm,width=3mm]},dashed] 
      (chor) -- (proj) -- ++ (2,0) |- (N);
\end{pgfonlayer}
\end{tikzpicture}
}
\caption{In choreographic programming, the collective behaviour of a system of processes is defined as a choreography, which is then automatically translated into a correct distributed implementation (drawing adapted from \cite{GMPRSW21}).}\label{fig:chor-prog}
\end{figure}

\paragraph*{The problem: communication failures}
Despite all the recent interest in choreographic programming and neighbouring approaches, like multiparty session types~\cite{HYC16}, the theories presented so far rely on strong assumptions about communication actions.
Most works just assume that communications never fail, which is clearly unrealistic in many scenarios for concurrent and distributed programming.
Other works allow for some communications or processes to fail, but with limitations. For example, they might assume synchronous communication to detect link failures~\cite{APN17}, that some processes are robust (never crash), that failures are permanent (crash, fail-stop), or reliable FIFO communications~\cite{VHEZ21}.
These limitations obscure the applicability of choreographic programming to real-world distributed systems in the future.

\paragraph*{This work}
In this work, we are interested in relaxing the assumptions of theory of choreographies all the way to the setting where every action related to communication can fail.
In terminology from the literature on failure models: processes can send or receive fewer messages than they are supposed to (\emph{send omission} and \emph{receive omission}, respectively)~\cite{H85,PT86}. Even when a send action succeeds in offloading a message to the network, the latter can still fail at delivering the message (\emph{omission failure})~\cite{C91}.
Our only assumptions are that messages do not get corrupted and participants are not malicious, i.e., they run the code projected from the choreography.
(This distinguishes our work from the most general setting of Byzantine failures, which we leave to future investigation.)

Weakening assumptions on communication in choreographies presents a novel challenge with two sides in tension. On the one hand, we need enough expressivity to program the local recovery strategies of processes, which in general may be asymmetric (e.g., exponential backoff at the sender and a timeout at the receiver).
This entails programming send and receive actions separately in choreographies.
On the other hand, we also need to retain the capability of expressing what communications the programmer wants to take place in choreographies, which is normally captured by atomic communication primitives \`a la $\proc p \to \proc q$.

We bridge the two aspects by deconstructing the choreographic term for communication in two elements: the declaration of intent to communicate, and its implementation as send and receive actions.
For example, the intent to communicate an integer from $\proc p$ to $\proc q$ is declared by writing
\begin{snippet*}
\init{k}{k'}{\proc p}{\proc q}{Int}
\end{snippet*}
which creates the \emph{frames} $k$ at $\proc p$ and $k'$ at $\proc q$ for communicating a value of type $\mathsf{Int}$(eger).
Frames are used as message tags by processes to implement communication actions (send or receive).
Continuing our example, $\proc p$ could use a procedure for exponential backoff to send a message, and $\proc q$ could use another procedure to attempt at receiving with a timeout.
\begin{snippet*}
\code{sendExpBackoff}(\proc p;k); \code{recvTimeout}(\proc q;k')
\end{snippet*}
This illustrates the asymmetry that we need to support: the policies for handling failures can be different for each process, and even each frame.

\paragraph*{Contributions}

We develop Lossy Choreographies (LC), the first language theory for choreographic programming that can deal with send omission, receive omission, and omission failures. 
These failures are formalised as rules for the operational semantics of \RC, which models fallible asynchronous communications.
Each type of omission is captured by a dedicated rule which allows to easily fine-tune the model to work with the desired failure mode.
Processes can deal with failures by performing local checks on their states.

LC offers a unified theory for different levels of abstraction: it is powerful enough to capture low-level recovery strategies based on retries under different conditions, but also to recover the usual simplicity of choreographic syntax through procedures and syntactic sugar.
For example, $\proc p.e \to^t_{\typef T} \proc q.x$ is sugar for an LC term where $\proc p$ sends a value of type $\typef T$ obtained by evaluating $e$ to process $\proc q$, and $\proc q$ either successfully receives this result in its variable $x$ or it times out after time $t$, in which case the value of $x$ does not get updated. LC thereby provides a foundation for choreographic languages where the standard way of writing a communication is with an arrow that is parametric on the recovery strategies the programmer wants used. This captures both the high-level abstraction of choreographic languages and the need for failure handling of the real world.
Our language can also capture more sophisticated protocols that combine different strategies.
This increase in expressive power extends the reach of choreographic programming to realistic transaction protocols based on voting for the first time -- we tested this by modelling two-phase commit~\cite{BN09} (the code is given in~\cref{sec:2pc}).

\looseness=-1
We develop a provably-correct notion of endpoint projection for LC, which translates choreographies into distributed code given in terms of a process calculus.
Here the key innovation is that communication declarations, which pair frames at sender and receiver at the choreographic level, can be implemented without performing synchronisation. 
Instead, we can achieve a fully distributed implementation by manipulating local counters that are guaranteed to remain in agreement because they were translated from the same choreography.
Thus, in a twist of fate, choreographies are the solution to their own problem regarding failures: the global view in choreographies provides a way for the local states of processes to agree by construction on how to correlate messages as intended by the programmer.
This elicits how choreographic programming benefits from its global view even when dealing with low-level details of communication implementation such as those studied here.

LC is carefully designed to enable static reasoning. In the declaration of a communication, the frame at the sender can be seen as a promise~\cite{FW78} to produce a value and the frame at the receiver as a future~\cite{BH77} from which a value might eventually be obtained.
If there are no omission failures, fulfilling the promise results in completing the future through the network.
Frames are checked to be used consistently with their declarations by a type system.
For example, the sender process in a communication declaration can later use its frame only for sending values of the right type.
We also provide a more sophisticated analysis, called \emph{robustness}, to reason about delivery guarantees under different modes of failures.
Under our weakest assumptions (send, receive, and omission failures), robustness guarantees best effort (every receive has the possibility of being successful) and at-most-once delivery (a message is never successfully read more than once by the receiver).
If omission failures are excluded, then robustness guarantees the stronger property of exactly-once delivery.
Settings where omission failures are typically excluded include: Inter-Process Communication, for example based on unnamed pipes in POSIX systems, shared memory, or file-based messages; and distributed systems using middleware for reliable transmission, like TCP (assuming no connection resets) or persistent message queues (based on AMQP, Apache Kafka, etc.).

In addition to our results, we have further validated our design by implementing it as a library for Choral, a state-of-the-art choreographic programming language \cite{GMP24} that supports real-world programming \cite{LM23}.
Our implementation shows that our ideas support the development of practical high-level abstractions for failure-aware choreographic programming.

\paragraph{Structure of the paper}
We start by introducing our language for modelling Lossy Choreographies in \cref{sec:chor-model}. We then give some examples of protocols modelled in this language in \cref{sec:examples}. We describe two systems, typing and robustness, for reasoning about the correctness of Lossy Choreographies in \cref{sec:robust}. In \cref{sec:synthesis} we present a language for local processes and a notion of endpoint projection for compiling choreographies to processes. We then describe our Choral implementation of Lossy Choreographies in \cref{sec:implementation}. Finally, we compare our development to related work in \cref{sec:related} and conclude in \cref{sec:concl}.

\section{Lossy Choreographies}
\label{sec:chor-model}

\subsection{Syntax}
The language of Lossy Choreographies (LC) is defined by the following grammar.
{\small\begin{align*}
C &\Coloneqq I;C \mid \cnil  &&
	I \Coloneqq \ginit \mid \gsend \mid \grecv \mid \gassign  \mid \gcond \mid \gcall \\
	s &\Coloneqq l \mid e
	&& b \Coloneqq e \mid c! \mid c? \mid c? l \indent\indent
	c \Coloneqq k \mid (\proc p,m)
\end{align*}}%
A choreography $C$ is a sequence of instructions ending in $\cnil$.
An instruction, ranged over by $I$, can be a local computation, a communication action, a conditional, or a procedure call. 
Term $\ginit$ declares a communication of a message of type $T$ from $\proc p$ to $\proc q$, binding the \emph{frame names} $k$ (for the sender) and $k'$ (for the receiver) to the continuation.
At runtime, when a communication declaration is executed, frame names are substituted with pairs -- $(\proc q, m)$ -- that contain the name of the other process in the communication ($\proc q$) and a natural number $m$ (also called \emph{frame number}) that identifies this specific communication for these two processes.\footnote{Frame numbers are inspired by the sequence numbers found in TCP.}
Communications are then implemented with send and receive actions, given by terms of the form $\gsend$ (read `$\proc p$ sends $s$ on frame $c$') and $\recv{q}{c'}{x}$ (read `$\proc q$ receives a message on $c'$ and stores it in $x$'). These actions may fail, as we will show in our semantics.

Following standard practice from choreographic programming and session types, we syntactically distinguish when a process sends the result of a local expression ($e$) or a statically-defined value literal ($l$, also known as selection labels)~\cite{M23,HYC16}.
We leave the language of local expressions as a parameter of our theory, as in other choreographic languages~\cite{CHY12,HG22,M23,CMP23}.
Selection labels make the development of endpoint projection clearer, because it streamlines detecting whether all processes involved in a conditional communicate sufficient information about which branch has been chosen~\cite{CHY12}.
This standard solution adapts to our setting without major changes (we give the details in the appendix).

In the local assignment term $\gassign$, $\proc p$ assigns its local variable $x$ the value of local computation $s$. Since $s$ is a local computation located at $\proc p$, we do not annotate the variables in $s$ with their locations, and write $\proc p.x:= y$ rather than $\proc p.x:=\proc p.y$.
In a conditional $\gcond$, $\proc p$ evaluates the guard $b$ and chooses between the possible continuations $C_1$ and $C_2$ accordingly. We explain the meaning of each guard $b$ next:
\begin{itemize}
	\item $\proc p.e$: $\proc p$ chooses $C_1$ if evaluating $e$ yields 
$\literal{true}$, and $C_2$ otherwise; 
	
	\item $\proc p.c!$: $\proc p$ chooses $C_1$ if its last send attempt 
for $c$ was successful, and $C_2$ otherwise; 
	
	\item $\proc p.c?$: $\proc p$ chooses $C_1$ if its last receive 
attempt for $c$ was successful, and $C_2$ otherwise;
	
	\item $\proc p.c? l$: $\proc p$ chooses $C_1$ if it successfully
received the label $l$ on $c$, and $C_2$ otherwise.
\end{itemize}
Term $\gcall$ invokes procedure $X$ with arguments $\procs p$ and $\vec c$.
Procedures are defined by providing equations as usual in process calculi~\cite{SW01}.

\begin{example}\label{ex:21}
We formalise the procedures \code{sendExpBackoff} and \code{recvTimeout} from the introduction as a family of procedures indexed by the expression at the sender ($e$) and the variable at the receiver ($x$). We use some auxiliary variables to store the number of send attempts at the sender ($n$) and the timeout at the receiver ($to$).
\begin{snippet}
	\procdef{\code{sendExpBackoff}_{e}}{
		\proc p;k
	}{	%
			\\\indent
			\send{\proc p}{k}{e}; \commentline{Try sending}
			\\\indent
		\cond{\neg \proc p.k!}{ \commentline{Check whether send has succeeded}
      \\\indent\indent
      \assign{p}{n}{\code{waitAndReturn}(n)}; \commentline{Wait $2^n$ milliseconds and return $n+1$}
			\\\indent\indent
			\code{sendExpBackoff}_{e}\langle\proc p;k\rangle \commentline{Call procedure recursively}
			\\\indent\mspace{-10.0mu}
		}
		{ \cnil \commentline{If send is successful, finish}}
	}
\end{snippet}

\begin{snippet}
	\procdef{\code{recvTimeout}_{x}}{
		\proc p;k
	}{
			\\\indent
			\assign{\proc p}{now}{\code{currentTime()}};
			\\\indent
			\recv{\proc p}{k}{x}; \commentline{Try receiving}
			\\\indent
			\assign{\proc p}{to}{to - (\code{currentTime()} - now)};\\\indent
		\cond{\proc p.to > 0 \land \neg \proc p.k?}{ \commentline{Check whether receive has succeeded or time has elapsed}
			\\\indent\indent
			\code{recvTimeout}_{x}\langle p;k\rangle \commentline{Call procedure recursively}
			\\\indent\mspace{-10.0mu}
		}
		{ \cnil \commentline{If receive is successful or the time has elapsed, finish}}
	}
\end{snippet}
Based on these procedures, we can define the syntactic sugar shown in the introduction.
\begin{snippet}
\proc p.e \to^t_{\typef T} \proc q.x:T \defeq
\\
\indent
\ginit;\
\assign{\proc p}{n}{0};\
\code{sendExpBackoff}_{e}\langle \proc p; k\rangle;\
\assign{\proc q}{to}{t};\
\code{recvTimeout}_{x}\langle\proc q; k'\rangle;\cnil
\end{snippet}
\end{example}

LC is also expressive enough to capture other strategies as syntactic sugar, including acknowledgements and compensations (custom code that is triggered in case of failure). (Examples are given in~\cref{sec:examples}.)

\subsection{Semantics}

\begin{figure}[t]\small
\begin{infrules}
  \infrule[\rname[C]{Assign}][rule:c-assign]{
    \tuple{\gassign;C,\Sigma,K}
    \lto{\tau @\proc p}
    \tuple{C,\Sigma[\proc p.x\mapsto v],K}
  }{
    \eval s\Sigma pv
  }
	\infrule[\rname[C]{Frame}][rule:c-frame]{
    \tuple{\gnewframe;C,\Sigma,K}
    \lto{\tau @\proc p}
    \tuple{C[(\proc q,n)/k],\Sigma',K}
  }{
    \Sigma(\proc p.\fcnt.\proc q)=n 
    \quad
    \Sigma' = \Sigma[
    \proc p.\fcnt.\proc q\mapsto n+1][\proc p.\fbuf.\proc q.n\mapsto \bot]
  }
  \infrule[\rname[C]{Com}][rule:c-frame-pair]{
    \tuple{\gnewframepair;C,\Sigma,K}
    \lto{\mu}
    \tuple{C',\Sigma',K}
  }{
    \tuple{\gnewframe;\newframe{k'}{q}{p}{T};C,\Sigma,K}
    \lto{\mu}
    \tuple{C',\Sigma',K}
		\quad
		\mu \in \{\tau @\proc p,\tau @\proc q\}
  }
  \infrule[\rname[C]{Send}][rule:c-send]{
    \tuple{\gssend;C,\Sigma,K}
    \lto{\proc p.(\proc q,m) \rightarrow}
    \tuple{C,\Sigma[\proc p.\fbuf.\proc q.m\mapsto \checkmark],K\uplus (\proc p,\proc q,m,v)}
  }{
    \eval s\Sigma p v
  }
  \infrule[\rname[C]{SendFail}][rule:c-send-fail]{
    \tuple{\gssend;C,\Sigma,K}
    \lto{\proc p.(\proc q,m) \rightarrow}
    \tuple{C,\Sigma,K}
  }{
    \eval s\Sigma pv
  }
  \infrule[\rname[C]{Del}][rule:c-deliver]{
    \tuple{C,\Sigma,K\uplus (\proc p,\proc q,m,v)}
    \lto{\tau}
    \tuple{C,\Sigma[\proc q.\fbuf.\proc p.m\mapsto v],K}
  }{
  }
  \infrule[\rname[C]{Loss}][rule:c-loss]{
    \tuple{C,\Sigma,K\uplus (\proc p,\proc q,m,v)}
    \lto{\tau}
    \tuple{C,\Sigma,K}
  }{
  }
  \infrule[\rname[C]{Recv}][rule:c-recv]{
    \tuple{\gsrecv;C,\Sigma,K}
    \lto{\rightarrow \proc p.(\proc q,m)}
    \tuple{C,\Sigma[\proc p.\fbuf.\proc q.m\mapsto v \checkmark][\proc p.x\mapsto v],K}
  }{
    \Sigma(\proc p.\fbuf.\proc q.m)\in\{v,v\checkmark\}
  }
  \infrule[\rname[C]{RecvFail}][rule:c-recv-fail]{
    \tuple{\gsrecv;C,\Sigma,K}
    \lto{\rightarrow \proc p.(\proc q,m)}
    \tuple{C,\Sigma,K}
  }{
    \Sigma(\proc p.\fbuf.\proc q.m)=\bot
  }
  \infrule[\rname[C]{Then}][rule:c-then]{
    \tuple{\gcond;C,\Sigma,K}
    \lto{\mathsf{left}@\proc p}
    \tuple{C_1\fatsemi C,\Sigma,K}
  }{\begin{array}{ll}  
    \eval b\Sigma p\true
  \end{array}
  }
  \infrule[\rname[C]{Else}][rule:c-else]{
    \tuple{\gcond;C,\Sigma,K}
    \lto{\mathsf{right}@\proc p}
    \tuple{C_2\fatsemi C,\Sigma,K}
  }{\begin{array}{ll}  
    \eval b\Sigma p\false
  \end{array}
  }
  \infrule[\rname[C]{DelayI}][rule:c-delay-i]{
    \tuple{I;C,\Sigma,K} \lto{\mu} \tuple{I;C',\Sigma',K'}
  }{
    \tuple{C,\Sigma,K} \lto{\mu} \tuple{C',\Sigma',K'}
    \and \pn(I) \cap \pn(\mu) = \emptyset
  }
  \infrule[\rname[C]{DelayC}][rule:c-delay-c]{
    \tuple{\gcond;C,\Sigma,K} \lto{\mu} \tuple{\cond{\proc p.b}{C_1'}{C_2'};C,\Sigma',K'}
  }{
    \tuple{C_1,\Sigma,K} \lto{\mu} \tuple{C'_1,\Sigma',K'}
    \and \tuple{C_2,\Sigma,K} \lto{\mu} \tuple{C'_2,\Sigma',K'}
    \and \proc p \notin \pn(\mu)
  }
  \infrule[\rname[C]{Call}][rule:c-call]{
    \tuple{\gcall;C',\Sigma,K}
    \lto{\tau@\proc r}
    \tuple{\procs{p}\setminus \proc r:\cont{X}{\procs{p};\vec{c}}.C';C[\procs{p}/\procs{ q}][\vec{c}/\vec{k}]\fatsemi C',\Sigma,K}
  }{
    \gprocdef \in \cdefs
    \qquad
    \proc r \in \procs p
  }
  \infrule[\rname[C]{Enter}][rule:c-enter]{
    \tuple{\gcont;C,\Sigma,K}
    \lto{\tau@\proc r}
    \tuple{\procs{q}\setminus \proc r:\cont{X}{\procs{p};\vec{c}}.C';C,\Sigma,K}
  }{
    \proc r \in \procs q \and \procs q\setminus\proc r\neq\emptyset
  }
  \infrule[\rname[C]{Finish}][rule:c-finish]{
    \tuple{\proc{r}:\cont{X}{\procs{p};\vec{c}}.C';C,\Sigma,K}
    \lto{\tau@\proc r}
    \tuple{C,\Sigma,K}
  }{}
\end{infrules}
\caption{Semantics of Lossy Choreographies}\label{fig:sem}
\end{figure}

We define an operational semantics for LC in terms of a labelled transition system (LTS), whose rules are given in \cref{fig:sem}.
States in this system are triples of the form $\langle C,\Sigma,K\rangle$ called \emph{configurations}.
We describe the additional elements $\Sigma$ and $K$ before proceeding to an explanation of the rules.
In the remainder, we write $\pn$ for function that returns the set of process names that appear in a choreography or transition label.
\begin{description}
	\item[$\Sigma$]
	A \emph{choreographic store} $\Sigma$ is a map from process names to their local memory stores ranged over by $\sigma$, similarly to several previous theories~\cite{CHY12,JV22,CMP23}.
We write $\Sigma(\proc p.x) = v$ to denote that $\Sigma(\proc p) = \sigma$ such that $\sigma(x) = v$, read `variable $x$ at process $\proc p$ has value $v$'.
Differently from prior work, we reserve two locations ($\fcnt$,$\fbuf$) to store data that processes use for implementing frames.
The reserved location $\fcnt$ stores counters used by the current process to generate frame identifiers, one for each other process: $\Sigma(\proc p.\fcnt.\proc q)$ is the counter at $\proc p$ used for frames from/to $\proc q$.
The reserved location $\fbuf$ tracks the state of frames used by the current process. We write $\Sigma(\proc p.\fbuf.\proc q.n)$ for the state of the frame that $\proc p$ uses for sending or receiving a message to $\proc q$ with identifier $n$.
This state can take different values:
\begin{itemize}[nosep]
	\item it is $\bot$ when the frame is first created,
	\item it is $\checkmark$, when the frame is outgoing and the frame has been successfully handed over to the network,
	\item it is $v$ when the frame is incoming and the network has delivered the value $v$ for this frame but the process has not consumed it yet, and 
	\item it is $v \checkmark$ when the value has been finally read by the process.
\end{itemize}
Frame counters and identifiers are similar to the device used in TCP to correlate asynchronous messages at sender and receiver.
The idea is that:
\begin{itemize}
\item each process maintains a counter for each other process it interacts with;
\item frame declarations increment counters locally, i.e., without synchronising with the other party;
\item frames are assigned the value held by the corresponding counter when they are created.
\end{itemize}
Our semantics ensures if frame counters in a store are consistent with a choreography, they will remain so through its execution so it suffices for processes to start counting frames from a set value --- we formalise this property after we present the semantics (\cref{def:concistent-store}).
	\item[$K$] A \emph{communication transit} $K$ models the network: it is a multiset of messages that have been successfully handed over to the network from the sender, but they have not been delivered to the receiver yet. Each message has the form $(\proc p,\proc q,m,v)$ where $\proc p$ is the sender, $\proc q$ is the receiver, $m$ is the frame number for this message from $\proc p$ to $\proc q$, and and $v$ is the payload. Note that $K$ can contain multiple messages on the same frame, and that messages on the same frame do not necessarily have the same payload. This allows e.g. for a sender to send multiple messages with different timestamps included in their payload and have them all be on their way to the receiver at the same time. However, $K$ is a multiset so the model does not assume any ordering guarantee.
\end{description}

We now move on to the rules, where we use transition labels to denote a local action at a process ($\tau@\proc p$ for assignments, $\mathsf{left}@\proc p$ and $\mathsf{right}@\proc p$ for the evaluation of conditionals), an unobservable action by the network ($\tau$), the creation of a new frame ($\proc p \to \proc q$), a send attempt ($\proc p.(\proc q,m) \to$), or a receive attempt ($\to \proc p.(\proc q,m)$). Our rules are parametric on a set of procedure definitions, $\cdefs$ which is used for recursion. For example, $\cdefs$ may include the definitions of $\code{sendExpBackoff}_{e}(\proc p;k)$ and $\code{recvTimeout}_{x}(p;k)$.

\Cref{rule:c-assign}, for local assignments, is standard: the update notation $\Sigma[\proc p.x \mapsto v]$ denotes a choreographic store such that $(\Sigma[\proc p.x \mapsto v])(\proc p.x) = v$ and behaves like $\Sigma$ otherwise~\cite{M23}.
The premise $\eval s\Sigma pv$ states that the evaluation of 
$s$ in the process store $\Sigma(\proc p)$ yields value $v$ -- this relation for local evaluation is a parameter of our theory and can be nondeterministic. We assume that selection labels are always evaluated to themselves.

\Cref{rule:c-frame-pair,rule:c-frame} model the creation of a pair of frames for a communication from $\proc p$ to $\proc q$ without relying on any rendezvous mechanism.
Instead, these rules rely on the runtime term $\gnewframe$ to represent the creation of a frame to/from $\proc q$ at $\proc p$:
in \cref{rule:c-frame}, $\gnewframe$ is consumed, the frame $k$ is assigned a number $n$, its state is initialised to $\bot$, and the frame counter for $\proc q$ is incremented; and in \cref{rule:c-frame-pair}, the semantics of $\gnewframepair$ is essentially defined as that of $\gnewframe; \newframe{k'}{q}{p}{T}$.
The order of the runtime terms in the premise of \cref{rule:c-frame-pair} is immaterial since the semantics allows these to be executed in any order via \cref{rule:c-delay-i} (explained below).

\Cref{rule:c-send,rule:c-send-fail} capture the possible executions of a send attempt for a frame $(\proc q,m)$ at $\proc p$.
The first, \Cref{rule:c-send} models a successful send: a payload $v$ is computed and the message $(\proc p,\proc q,m,v)$ is successfully handed off to the network transit ($K$). It also updates the frame state at the sender accordingly.
\Cref{rule:c-send-fail}, instead, models a \emph{send omission} failure: the send action is consumed but it omits (fails at) adding the message to $K$.
In both cases, no information about the attempt is propagated to the receiver: our semantics is asynchronous and the only way to exchange information across processes is through fallible communication actions.
Note that \cref{rule:c-send} does not check the state of the frame, to reflect the real-world situation that a sender may perform multiple send actions resulting in the transmission of different payloads for the same frame.

Once a message is in transit, there are two possibilities:
It can either be successfully delivered to the receiver, which causes a corresponding update to its frame state (\cref{rule:c-deliver}), or incur an \emph{omission failure} and be lost (\cref{rule:c-loss}).
The sender has no knowledge of what happens.

\Cref{rule:c-recv,rule:c-recv-fail} define the execution of a 
receive attempt. The attempt can be successful (\ref{rule:c-recv}) only if the receiver's state for the desired frame contains a value (previously put there by \cref{rule:c-deliver}).
Otherwise, if no message for that frame has reached the receiver yet, the receive attempt fails (\cref{rule:c-recv-fail}).

\Cref{rule:c-then,rule:c-else} model conditionals. We formalise the intuition given for guards in the syntax by assuming the following local evaluation rules for any $\proc p$ and $\Sigma$:

\Cref{rule:c-call,,rule:c-enter,,rule:c-finish} unfolds procedure calls by replacing 
all occurrences of formal arguments in its body ($C$) with actual ones as prescribed by the 
substitution $\vec{\proc p};\vec{c}$ (provided its domain of definition coincides with the set $\vec{\proc q};\vec{k}$) and all 
names bound in the procedure body $C$ with fresh ones as per Barendregt's convention. To do this we use a runtime term, $\procs{q}:\cont{X}{\procs{p};\vec{c}}.C$, which denotes that $C$ is the choreography we had before calling procedure $X(\procs p;\vec{c})$, and of the participants $\procs p$ in $X$, $\procs q$ have yet to enter $X$. 

\begin{infrules}
\infrule{\eval {(\proc q,m)?}\Sigma p\true}{\Sigma(\proc p.\fbuf.\proc q.m)=v\checkmark}
\infrule{\eval {(\proc q,m)!}\Sigma p\true}{\Sigma(\proc p.\fbuf.\proc q.m)=\checkmark}
\infrule{\eval {(\proc q,m)?l}\Sigma p\true}{\Sigma(\proc p.\fbuf.\proc q.m)=l\checkmark}
\\
\infrule{\eval {(\proc q,m)?}\Sigma p\false}{\Sigma(\proc p.\fbuf.\proc q.m)\neq v\checkmark}
\infrule{\eval {(\proc q,m)!}\Sigma p\false}{\Sigma(\proc p.\fbuf.\proc q.m)\neq\checkmark}
\infrule{\eval {(\proc q,m)?l}\Sigma p\false}{\Sigma(\proc p.\fbuf.\proc q.m)\neq l\checkmark}
\end{infrules}

All other rules are standard for the theory of choreographic languages~\cite{M23}.
In the rules for conditionals \cref{rule:c-then,rule:c-else}, we use a meta-operator for sequential composition `$\fatsemi$'~\cite{CM17:forte,M23}. It replaces $\cnil$ in a choreography with another choreography:
\begin{equation*}
\cnil\fatsemi C \defeq  C
\qquad
(I;C)\fatsemi C' \defeq I;(C\fatsemi C').
\end{equation*}
Lastly, \cref{rule:c-delay-i,rule:c-delay-c} allow for executing actions performed by distinct processes out of order, as introduced in~\cite{CM13}. This reflects the fact that separate processes are executed concurrently. The operator $\disjoint$ checks that two sets are disjoint, and $\pn(C)$ and $\pn(\mu)$ denotes the process names mentioned in respectively a choreography and a label.

\def\frt#1#2{\phi^{\proc{#1}}_{\proc{#2}}}
\def\gfrt{\frt{p}{q}}

With the language fully defined, we formalise the property that frame states, counters, and runtime terms are consistent across a choreography and store.
\begin{definition}
	\label{def:concistent-store}
$\Sigma$ is \emph{consistent with} $C$ if, for any two $\proc p$ and $\proc q$ in $C$:
\begin{enumerate}[nosep]
	\item $\Sigma(\proc p.\fbuf.\proc q.n)$ is defined for any $n < \Sigma(\proc p.\fcnt. \proc q)$; and 
	\item $|\gfrt(C)| = \max(0, \Sigma(\proc p.\fcnt. \proc q)-\Sigma(\proc q.\fcnt. \proc p))$ where the function $\gfrt$ computes the set of frame names to $\proc q$ that $\proc p$ has yet to initialise as follows.
	\begin{gather*}
		\gfrt(\cnil) = \emptyset
		\qquad
		\gfrt(I; C) = \gfrt(I) \cup \gfrt(C) \text{ where } \gfrt(I) \cap \gfrt(C) = \emptyset
		\\
		\gfrt(\gcond) = \gfrt(C_1) \text{ where } {\gfrt(C_1) = \gfrt(C_2)}
		\\
		\gfrt(\gnewframe) = \{k\}
		\quad
		\gfrt(I) = \emptyset \text{ for } I \notin \{\gnewframe,\gcond\} 
	\end{gather*}
\end{enumerate}
\end{definition}
Note that the function $\gfrt$ is partial and rejects choreographies that misuse runtime terms for frame creation e.g., the condition on the case $I; C$ excludes repetitions of runtime terms for the same $k$.

\begin{proposition}
If $\tuple{C,\Sigma,K} \lto{\mu} \tuple{C',\Sigma',K'}$ and $\Sigma$ is consistent with $C$, then $\Sigma'$ is consistent with $C'$.
\end{proposition}

\subsection{Understanding failures}
\label{sec:understanding-failures}
We illustrate and explain the realism of our model wrt communication failures with a simple end-to-end communication.
Consider the following choreography, where $\proc p$ attempts to communicate the number $3$ to $\proc q$.
\begin{snippet*}
	 \send{p}{k}{3}; \;
	 \recv{q}{k'}{x}; \;
	 \cnil
\end{snippet*}

\begin{figure}[t]
\centering
\resizebox{\textwidth}{!}{
\begin{tikzpicture}[auto,yscale=1.6,xscale=3.7,
		state/.style={
			outer sep=0pt,
			inner sep=1pt,
		},
		transition/.style={
			->,
			rounded corners=5pt,
		},
		transition loss/.style={
			transition,
			densely dashed,
			red
		},
		transition sendfail/.style={
			transition,
			densely dashed,
			red
		},
		transition recvfail/.style={
			transition,
 			densely dashed,
			red
		},
		transition label/.style={
			font=\scriptsize,
		}
	]

	\newcounter{stc}
	\def\state(#1) at (#2) <#3|#4|#5>{
		\stepcounter{stc}
		\node[state] (#1) at (#2) {
			\(\left\langle\begin{matrix}
        \code{#3},\\\ensuremath{#4},\\\ensuremath{#5}
      \end{matrix}\right\rangle\)
		};
    \node[black!80,font=\scriptsize,anchor=north,xshift=-3pt] at (#1.west) {%
      \customlabel{state}{st:#1}{(\alph{stc})}%
      \hypertarget{st:#1}{{\textsf{(\alph{stc})}}}%
    };
	}

	\state (nB) at (1,2) <$\send{p}{(\proc q,1)}{3};\recv{q}{(\proc p,1)}{x};\cnil$|\Sigma|\emptyset>
	\state (nC) at (2,1) <$\recv{q}{(\proc p,1)}{x};\cnil$|\Sigma[\proc p.\fbuf.\proc q.1 \mapsto \checkmark] |\{(\proc p,\proc q,1,3)\}>
	\state (nD) at (3,2) <$\recv{q}{(\proc p,1)}{x};\cnil$|\Sigma'[\proc q.\fbuf.\proc p.1 \mapsto 3]|\emptyset>
	\state (nE) at (2,3) <$\cnil$|\Sigma'[\proc q.\fbuf.\proc p.1 \mapsto 3\checkmark][\proc q.x\mapsto 3]|\emptyset>
	\state (nF) at (1,1) <$\send{p}{(\proc q,1)}{3};\cnil$|\Sigma|\emptyset>
	\state (nG) at (0,1) <$\recv{q}{(\proc p,1)}{x};\cnil$|\Sigma|\emptyset>
	\state (nH) at (0,0) <$\cnil$|\Sigma|\emptyset>
	\state (nI) at (2,0) <$\cnil$|\Sigma[\proc p.\fbuf.\proc q.1\mapsto \checkmark] |\{(\proc p,\proc q,1,3)\}>
	\state (nJ) at (3,0) <$\recv{q}{(\proc p,1)}{x};\cnil$|\Sigma'|\emptyset>
	\state (nK) at (3,-1) <$\cnil$|\Sigma'|\emptyset>
	\state (nL) at (1,-1) <$\cnil$|\Sigma'[\proc q.\fbuf.\proc p.1\mapsto 3]|\emptyset>

	\draw[transition] (nB) --
		node[transition label] {\ref{rule:c-send}} 
		(nC);
	\draw[transition] (nC) --
		node[transition label] {\ref{rule:c-deliver}} 
		(nD);
	\draw[transition] (nD) --
		node[transition label] {\ref{rule:c-recv}} 
		(nE);
	\draw[transition recvfail] (nB) -- 
		node[transition label] {\ref{rule:c-recv-fail}} 
		(nF);
	\draw[transition sendfail] (nB) -- 
		node[transition label,swap] {\ref{rule:c-send-fail}} 
		(nG);
	\draw[transition sendfail] (nF) -- 
		node[transition label] {\ref{rule:c-send-fail}} 
		(nH);
	\draw[transition] (nF) --
		node[transition label,swap] {\ref{rule:c-send}} 
		(nI);
	\draw[transition recvfail] (nG) -- 
		node[transition label,swap] {\ref{rule:c-recv-fail}} 
		(nH);
	\draw[transition recvfail] (nC) -- 
		node[transition label] {\ref{rule:c-recv-fail}} 
		(nI);
	\draw[transition loss] (nC) -- 
		node[transition label] {\ref{rule:c-loss}} 
		(nJ);
	\draw[transition] (nI) --
		node[transition label] {\ref{rule:c-deliver}} 
		(nL);
	\draw[transition loss] (nI) -- 
		node[transition label, swap] {\ref{rule:c-loss}} 
		(nK);
	\draw[transition recvfail] (nJ) -- 
		node[transition label] {\ref{rule:c-recv-fail}} 
		(nK);
    
\node[anchor=north] at (1.5,-1.5) {\(
\Sigma(\proc p.\fbuf.\proc q.1)=\Sigma(\proc p.\fbuf.\proc q.1)=\bot \qquad \Sigma' = \Sigma[\proc p.\fbuf.\proc q.1 \mapsto \checkmark]\)};
\end{tikzpicture}
} %
\caption{Labelled transition system illustrating all possible executions of an end-to-end communication in LC -- dashed (red) arrows are transitions that model failures.}
\label{fig:com-execution}
\end{figure}

The LTS in \cref{fig:com-execution} captures all the possible executions for this program.
For convenience of exposition, we assign a letter to each state. Also, we do not show transition labels but rather the name of the rule applied to derive the transition -- the transition from \ref{st:nB} to \ref{st:nF} is the only one that also requires \cref{rule:c-delay-i}.

The transition system begins in state \ref{st:nB} and every run eventually terminates (reaches the choreography $\cnil$).
We describe the different situations at the ends of the possible executions.
\begin{description}
\item[\ref{st:nE}] This is the final state of the only successful execution, which requires no send, receive, or omission failures.
The value $3$ is marked as delivered and no transition from \ref{st:nB} to \ref{st:nE} uses any of the rules that model failures.

\item[\ref{st:nH}] This configuration is reached only if we have both a send and a receive omission.
There are two paths to \ref{st:nH}, one passing through \ref{st:nG} and one through \ref{st:nF}.
In the former, a send omission stops the receive from ever succeeding. In the latter, we have a receive omission because the receive is attempted before the send action had a chance to put the message in the network. The latter case exemplifies how our semantics captures timeouts at the receiver. In fact, \ref{st:nF} may also transition to \ref{st:nI}: this is the case where the receive fails purely because of unfortunate timing, i.e., it takes place before the successful send.

\item[\ref{st:nK}] This state is reached because of an omission failure. In the path through \ref{st:nF}, the receive fails regardless of the omission failure because of timing. In the path through \ref{st:nJ}, the omission failure is the cause for the receive to fail.

\item[\ref{st:nL}] This state is reached if the send succeeds but the receive fails because of timing issues: in the path through \ref{st:nF}, the receive is executed before the send; in the path through \ref{st:nC}, the receive is executed before the network could carry the message to the receiver.

\end{description}

\section{Applications}
\label{sec:examples}
We now give some examples of the kinds of protocols LC can model.
\subsection{Higher-level communication patterns}
\label{sec:higher-level-com}

\paragraph*{Acknowledgements}
Assume a setting with omission failure.
In this setting, the definition of $\proc p.e \to^t_{\typef T} \proc q.x:T$ presented in \cref{ex:21} risks ending up in a state where the sender believes it has completed a communication but the receiver has timed out due to the message being lost in the network or the send taking too long.

This is a common problem in practice, which is addressed by switching to ``best-effort'' strategies where delivery is possible (to varying degrees) but not certain.
Below is a procedure that implements a simple communication protocol with capped retries and acknowledgements to the sender. In this, the strategy implemented by $\code{comACK}_{e,x}$ can be regarded as a simplification of that of TCP; four-phase handshakes or other protocols are implementable in LC as well.
\begin{snippet}
	\procdef{\code{comACK}_{e,x}}{
		\proc s,
		\proc r;\emptyset
	}{
		\\\indent
		\init{k}{k'}{s}{r}{T}; \commentline{create new frame for communication}
		\\\indent
		\init{k'_{ack}}{k_{ack}}{r}{s}{Unit}; \commentline{Create new frame for acknowledgement}
		\\\indent
		\code{sendExpBackoff}_e(\proc s;k)\conc \commentline{Send}
		\\\indent
		\code{recvTimeout}_x(\proc r;k')\conc \commentline{Receive}
		\\\indent
		\code{sendExpBackoff}_{unit}(\proc r;k_{ack}')\conc \commentline{Send acknowledgement}
		\\\indent
		\code{sendUntilACK}_{e,x}(\proc s;k,k_{ack})\conc \commentline{Call send procedure}
	}
	\\
	\procdef{\code{sendUntilACK}_{e,x}}{
		\proc s;
		k,
		k_{ack}
	}{ \\\indent
		\code{sendExpBackoff}_e(\proc s;k)\conc \commentline{Send}
		\\\indent
		\code{recvTimeout}_x(\proc s;k_{ack})\conc \commentline{Try receiving acknowledgement}
		\\\indent
		\cond{\proc s.(n > 0) \wedge \neg \proc s.k_{ack}?}{ \commentline{Check number of attempts and received acknowledgement}
			\\\indent\indent
		\proc s.n:=n-1; \commentline{Update send attempt number}
			\\\indent\indent
			\code{sendUntilACK}_{e,x}(\proc s;k,k_{ack})\conc \commentline{Recursive call}
			\\\indent\mspace{-10.0mu}
		}
		{\cnil \commentline{When acknowledgement has been received or we run out of attempts, end.}}
	}
\end{snippet}

\paragraph*{Compensations}
With $\code{comACK}_{e,x}$ we can also use LC to develop a new variant of choreographies that does not assume reliable transmission, i.e., when we are in a setting with omission failures. 
In this setting, a common pattern to deal with failures of best-effort communications are 
\emph{compensations}. Fault compensations can be defined in LC (for both 
settings with and without omission failures) using conditionals, $\code{comACK}_{e,x}$ (or variations thereof), and 
some syntax sugar to improve readability.
An expression ${\proc s.e \Rightarrow^{BE} \proc r.x}\{C_{\proc s}\}\{C_{\proc r}\}$ is a communication as 
in $\code{comACK}_{e,x}(\proc s,\proc r;\emptyset)$ where choreographies $C_{\proc s}$ and $C_{\proc r}$ are 
executed as compensations for faults detected by the sender $\proc s$ (no ack) or the receiver $\proc 
r$, respectively.
An example of communications with fault compensations is the communication construct defined in 
\cite{APN17} where communication operations specify default values as compensations; this
is recovered in LC using local computations as, e.g., in ${\proc s.e \Rightarrow^{BE} \proc r.x}\{\proc 
s.x:=\literal{foo}\}\{\proc r.x:=\literal{42}\}$.

\paragraph*{Any/Many communications}
We can also implement more complex communication primitives, like those in \cite{LNN16,CMP18}.
Below are procedures that iteratively attempt
at sending some frames until the sender 
stack accepts all or any of them, respectively, using a round-robin strategy.
\begin{snippet}
	\procdef{\code{sendAll}_{n,e}}{
		\proc s;
		k_1,
		\dots,
		k_n
	}{\\\indent
		\send{\proc s}{k_n}{e}; \commentline{Send to the last frame in the list}
		\\\indent
		\cond{\proc s.k_n!}{ \commentline{Check if send was successful} %
		\\\indent\indent
				\code{sendAll}_{n-1,e}(
					\proc s;
					k_1,
					\dots,
					k_{n-1}) \commentline{Send to the remaining frames}
					\\\indent
		}{ 
		\\\indent\indent
			\code{sendAll}_{n,e}(\proc s;
				k_2,
				\dots,
				k_n,
				k_1)  \commentline{Otherwise, try this frame again later}
		} %
	}
\end{snippet}
\begin{snippet}
	\procdef{\code{sendAny}_{n,e}}{
		\proc s;
		k_1,
		\dots,
		k_n,
	}{\\\indent
		\send{\proc s}{k_1}{e}; \commentline{Send to the first frame in the list}
		\\\indent
		\cond{\neg\proc s.k_1!}{ \commentline{Check if send was unsuccessful}
		\\\indent\indent
			\code{sendAny}_{n,e}(\proc s;
				k_2,
				\dots,
				k_n,
				k_1) 
        \commentline{Retry cycling through the frames}
		\\\indent}
		{
		 \cnil \commentline{End function if the send succeeded}}
	}
\end{snippet}
We omit the dual procedures for receiving all or some frames, which are similarly defined.
Combining these it is possible to implement scatter/gather communication primitives from 
\cite{LNN16}. For instance, below is an implementation of scatter.
\begin{snippet}
	\procdef{\code{scatterAll}_{e,x}}{
		\proc s,
		\proc r_1,
		\dots,
		\proc r_n;\emptyset
	}{\\\indent
		{\init{k_1}{k'_1}{\proc s}{\proc r_1}{T};
		\dots
		\init{k_n}{k'_n}{\proc s}{\proc r_n}{T}}; \commentline{Create frames}
		\\\indent
		\code{sendAll}_e(\proc s;k_1,\dots,k_n); \commentline{Call send all on the new frames}
		\\\indent
		\code{recvTimeout}_x(\proc r_1;k_1');
		\dots
		\code{recvTimeout}_x(\proc r_n;k_n') \commentline{Each receiver receives on its frame.}
	}
\end{snippet}
\subsection{Two-phase commit}\label{sec:2pc}
We now show that LC can be used to implement common protocols for dealing with failures. The two-phase commit protocol \cite{BN09} is a protocol for getting a set of participants to agree to either all commit or all abort a transaction in a setting with unreliable communication. 

All the participants send a vote to the controller telling it whether they want to commit. If the controller receives yes votes from all participants, then it decides to go through with the commit. Otherwise if any votes are no or are not successfully communicated to the controller, it decides to abort.
The controller then sends its decision to each participant, trying repeatedly to send until it receives an acknowledgement. 
Meanwhile the participants all wait to receive the decision, and when they receive it perform the abort or commit and repeatedly send the acknowledgement until the send is successful.
We base our code on the version of the two phase commit protocol presented in \cite{BN09}.
\begin{snippet}
	\procdef{\code{2PhaseCommit}}{
		\proc c,
		\proc {p_1},
		\dots,
		\proc {p_n};
	}{
	\\\indent
		{\init{k_{v_1}}{k'_{v_1}}{\proc {p_1}}{\proc c}{Bool};
		\dots;
		\init{k_{v_n}}{k'_{v_n}}{\proc {p_n}}{\proc c}{Bool};} \commentline{Frames for votes}
		\\\indent
		{\init{k_{d_1}}{k'_{d_1}}{\proc c}{\proc {p_1}}{Bool}; 
		\dots;
		\init{k_{d_n}}{k'_{d_n}}{\proc c}{\proc {p_n}}{Bool};} \commentline{Frames for decision}
		\\\indent
		{\init{k_{a_1}}{k'_{a_1}}{\proc {p_1}}{\proc c}{Unit};
		\dots;
		\init{k_{a_n}}{k'_{a_n}}{\proc {p_n}}{\proc c}{Unit};} \commentline{Frames for acknowledgements}
		\\\indent
		{\assign{c}{v_1}{false};
		\dots;
		\assign{c}{v_n}{false};} \commentline{All votes start out false at $\proc c$}
		\\\indent
		{\send{p_1}{k_{v_1}}{\mathsf{vote}()};
		\dots;
		\send{p_n}{k_{v_n}}{\mathsf{vote}()};} \commentline{Participants send votes}
		\\\indent
		{\recv{c}{k'_{v_1}}{v_1};
		\dots;
		\recv{c}{k'_{v_n}}{v_n};} \commentline{Controller receives votes (they remain false if receive fails)}
		\\\indent
		\assign{c}{decision}{v_1\wedge \dots \wedge v_n}; \commentline{Decision on whether to commit is made - only if all voted yes}
		\\\indent
		\code{sendAllUntilAck}(\proc c;k_{d_1},\dots,k_{d_n},k'_{a_1},\dots,k'_{a_n}); \commentline{$\proc c$ calls a procedure to send the decision}
		\\\indent
		\code{RecvDec}(\proc {p_1};k'_{d_1},k_{a_1});
		\dots;	
		\code{RecvDec}(\proc {p_n};k'_{d_n},k_{a_n}) \commentline{$\proc p$s call procedures to receive decision}
	}
	\\
	\procdef{\code{sendAllUntilAck}_n}{\proc c;
	 k_{d1}, \dots, k_{d_n},k_{a_1},\dots,k_{a_n}}
	{
	\\\indent
	\send{c}{k_{d_n}}{decision}; \commentline{$\proc c$ sends the decision to a participant}
	\\\indent
	\recv{c}{k_{a_n}}{x}; \commentline{$\proc c$ receives acknowledgement from that participant}
	\\\indent
	\cond{k_{a_n}?}{ \commentline{If the acknowledgement was received}
	\\\indent\indent
	\code{sendAllUntilAck}_{n-1}(\proc c;
	 k_{d_1}, \dots, k_{d_{n-1}},k_{a_1},\dots,k_{a_{n-1}}) \commentline{Send to other participants}
	 \\\indent
	}{ 
	\\\indent\indent
	\code{sendAllUntilAck}_{n}(\proc c;
	 k_{d_2}, \dots, k_{d_n},k_{d_1},k_{a_2},\dots,k_{a_n},k_{a_1}) \commentline{Try again later}
	}
	}
	\\
	\procdef{\code{RecvDec}}{\proc p;k_{d},k_{a}}
	{
	\\\indent
	\recv{\proc p}{k_{d}}{dec}; \commentline{Try receiving the decision}
	\\\indent
	\cond{\proc p.k_{d}?}{ \commentline{If receive was successful}
	\\\indent\indent
	\cond{\proc p.dec}{ \commentline{Execute the decision to abort or commit}
	\\\indent\indent\indent
	\assign{\proc p}{memory}{\mathsf{commit}()}
	\\\indent\indent}
	{\\\indent\indent\indent
	\assign{\proc p}{memory}{\mathsf{abort}()}
	}
	\\\indent\indent
	\code{sendExpBackoff}_{unit}(\proc p,k_{a})  \commentline{Keep sending acknowledgement until successful}
	\\\indent}
	{
	\\\indent\indent
	\code{RecvDec}(\proc p;k_{d},k_{a}) \commentline{If receive decision was unsuccessful, try again}
	}
	}
\end{snippet}

Note that \code{sendAllUntilAck} is a combination of \code{sendAll} and \code{sendUntilAck}. For simplicity, we have removed the check on number of send attempts that we saw in \code{sendUntilAck}, but this could easily be reintroduced if one is concerned about $c$ being blocked by a lost acknowledgement.

\section{Reasoning about Lossy Choreographies}\label{sec:robust}
\subsection{Typing}\label{sec:types}
LC programs can get stuck if procedures are called on wrong arguments, communication actions are performed against the wrong frames or processes, and guards of conditionals are not of the expected type. 
In this section we introduce a type system for LC that rejects this kind of programs and ensures progress. 

Our type judgements are of the form $\Gamma\vdash C$ where $\Gamma$ an environment consisting of the local environments of processes and the frame types associated with predefined procedures. \[
\Gamma \coloneqq \call{X}{\vec{\proc p}}{\vec{k:\typef{T}}}
\mid \proc p.c \colon F 
\mid \proc p.x \colon \typef{T}
\qquad
F \coloneqq \sndt{T} \mid \rcvt{T}
\] The local environments of processes are accessed by using $\Gamma$ as a function such that $\Gamma(\proc p)=\gamma$ with $\gamma$ being a local environment mapping variables and frames to types $\gamma(x)=\typef{T}$ and $\gamma(c)\in\{\sndt{T},\rcvt{T}\}$ for local types $\typef{T}$. Here $\sndt{T}$ denotes a frame which sends a payload of type $\typef{T}$ and $\rcvt{T}$ a frame that receives a payload of type $\typef{T}$. Just like we assumed a user-defined local language, we do the same for the local type system, which we assume has local judgements $\gamma\vdash s:\typef{T}$ which include the local boolean judgements in \cref{fig:loctyp}. 

\begin{figure}[t]
\begin{infrules}
\infrule[\rname [B]{Sent}]{\Gamma,c:\sndt{T}\vdash c!:\typef{Bool}}{} 
\infrule[\rname [B]{Rcvd}]{\Gamma,c:\rcvt{T} \vdash c?:\typef{Bool}}{} 
\infrule[\rname [B]{Rcvd}]{\Gamma,c:\rcvt{\typef{Label}} \vdash c? l:\typef{Bool}}{}
\end{infrules}
\caption{Local boolean typing}\label{fig:loctyp}
\end{figure}

We show the typing rules for choreographies in \cref{fig:typ}. We use $\Gamma(\proc p.x)$ as shorthand for $\Gamma(\proc p)(x)$. 
\begin{figure}[t]
\begin{infrules}
	\infrule[\rname [T]{Com}][rule:t-frame-pair]
		{\Gamma\vdash \gnewframepair;C}
		{\Gamma,\proc p.k:\sndt{T},\proc q.k':\rcvt{T}  \vdash C}
	\infrule[\rname [T]{Frame}][rule:t-frame]
		{\Gamma\vdash \gnewframe;C}
		{\Gamma(\proc p.k) \in \{\sndt{T},\rcvt{T}\}
			\and \Gamma\vdash C} 
\infrule[\rname [T]{Assign}][rule:t-assign]{\Gamma\vdash \gassign;C}{\Gamma(\proc p)\vdash s:\typef{T} \and \Gamma(\proc p.x)=\typef{T} \and \Gamma\vdash C} 
\infrule[\rname[T]{Send}][rule:t-send]{\Gamma \vdash \gsend;C}{\Gamma(\proc p.c)=\sndt{T} \and \Gamma(\proc p)\vdash s:\typef{T} \and \Gamma \vdash C} 
\infrule[\rname [T]{Recv}][rule:t-recv]{\Gamma \vdash \grecv;C}{\Gamma(\proc p.c)=\rcvt{T}\and \Gamma(\proc p.x)=\typef{T} \and \Gamma \vdash C} 
\infrule[\rname [T]{Cont}][rule:t-cont]{\Gamma\vdash \gcont;C}{\Gamma \vdash C}
\infrule[\rname [T]{Nil}][rule:t-nil]{\Gamma \vdash \cnil}{} 
\infrule[\rname [T]{Cond}][rule:t-cond]{\Gamma\vdash \gcond}{\Gamma(\proc p)\vdash b:\typef{Bool} \and \Gamma\vdash C_1 \and \Gamma\vdash C_2} 
\infrule[\rname [T]{Call}][rule:t-call]{\Gamma,\call{X}{\vec{\proc q}}{\vec{k:\typef{T}}}\vdash \gcall;C}{\Gamma\vdash C \and \forall k_i,\typef{T}_i.\exists \proc {p_j}. \Gamma(\proc {p_j}.c_i)=\typef{T}_i}
\end{infrules}
\caption{Typing}\label{fig:typ}
\end{figure}

For the most part, these rules are fairly intuitive. \Cref{rule:t-frame-pair} adds both ends of the frame to the environment with the expected send and receive types and \cref{rule:t-frame} checks that the runtime term for frame creation refers to a frame in the environment.\footnote{For conciseness, neither rule check for misuses of runtime terms like $\gnewframe; \gnewframe$ or $\gnewframepair; \gnewframe$ as these are already excluded by \cref{def:concistent-store}; these checks amount to adding the premises $k \notin \frt{p}{q}(C)$ and $k' \notin \frt{q}{p}(C)$.} \Cref{rule:t-assign} checks that the type of the variable being updated is the same as its new value. \Cref{rule:t-send,rule:t-recv} check that the type of the frame matches the type of the payload and the variable it gets stored on. \Cref{rule:t-cont} does nothing because $\gcont$ is a runtime term that only serves to ensure every process enters a procedure before executing it. \Cref{rule:t-nil} also does nothing since $\cnil$ can be typed with any environment. \Cref{rule:t-cond} checks that the condition is a boolean and both branches can be typed. And finally, \cref{rule:t-call} checks that the frames used in the call of a procedure match the types of the frames that procedure uses in $\Gamma$. For this to make sense, we also use $\Gamma$ to type the set of procedure to ensure the definition of the procedure also uses frames with those types.
\begin{infrules}
\infrule[\rname [T]{Defs}][rule:t-defs]{\Gamma\vdash \cdefs}{
\cdefs(X(\vec{\proc q};\vec{k}))=C\wedge \Gamma(X(\vec{\proc q};\vec{k:T})) \Rightarrow \Gamma\vdash C \wedge \forall \proc{q_i},k_j,T_j.\Gamma(\proc{q_i}.k_j)\in\{T_j,\bot\}
}
\end{infrules}

We now give the definition of a well-types configuration. 
This definition ensures that messages in transit have values consistent with the types of the sending and receiving frames. 
\begin{definition}
We say that a configuration $\tuple{C,\Sigma,K}$ is \emph{well-typed} if there is a type environment $\Gamma$ such that 
\begin{itemize}[nosep]
\item $\Gamma\vdash C$; 
\item $\Gamma \vdash \Sigma(\proc p.x)\colon \Gamma(\proc p.x)$ for any variable $\proc p.x\in\Sigma$;
\item $\Gamma(\proc p.(\proc q,m))=\sndt{T}$,
$\Gamma(\proc q.(\proc p,m))=\rcvt{T}$, and 
$\vdash v\colon \typef{T}$
for any $(\proc p,\proc q,m,v)\in K$;
\item $\Sigma$ is consistent with $C$.
\end{itemize}
\end{definition}

We can now describe the standard properties of choreographies which are guaranteed by our type system.

First, we show typability preservation. If we can type a choreography then we can type any other choreography reachable from that.
\begin{proposition}[Typability Preservation]
	Given a well-typed configuration $\tuple{C,\Sigma,K}$ and set of procedure definitions $\cdefs$, if  $\tuple{C,\Sigma,K}\lto{\mu}\tuple{C',\Sigma',K'}$, then $\tuple{C',\Sigma',K'}$ is also well-typed.
\end{proposition}
\begin{proof}
Follows from case analysis on the transition. 
\end{proof}

Next we prove progress. Any typable state of a choreography either can do an action or has completed its computation. In most choreographic languages, this is one of the key properties guaranteed for choreographies, but as our communication actions are non-blocking, the only requirement for progress in this setting is that conditions evaluate to either true or false.
\begin{theorem}[Progress]\label{thm:progress}
	Given a well-typed configuration $\tuple{C,\Sigma,K}$ and set of procedure definitions $\cdefs$, $C = \cnil$ or $\tuple{C,\Sigma,K}\lto{\mu}\tuple{C',\Sigma',K'}$.
\end{theorem}
\begin{proof}
This property is simpler to prove than in most choreographic languages, since communication is asymmetric and receive fails instead of waiting for the sender. Therefore the only requirement to ensure progress is that a condition in a conditional is either true or false, which we ensure with our typing rules.
\end{proof}
\subsection{Robustness}
The last component of our calculus is an inference system for reasoning about \emph{robustness}, i.e., what delivery guarantees a choreography provides in the presence of arbitrary failures. Our robustness system checks that 
\begin{itemize}
\item Communication is at most once and best effort;
\item There are no unnecessary checks on network actions (to avoid dead branches).
\end{itemize}
Robustness judgements have the form $\rbst{P}{R}{C}{F_0}{F_1}$. Here $C$ is a choreography; $P$ is a list of pairs of connected frames (by default $((\proc p.(\proc q,m),\proc q.(\proc p,m))$ for all processes $\proc p,\proc q$ in the choreography and all $m$; and $F_0$ and $F_1$ are abstract frame dictionaries before and after the execution of $C$. An abstract frame dictionary is a list of frames and potential payloads of the form $\proc p.c:U$ where $U$ is a set consisting of $\curlywedge$, $\bullet$, and any label $l\in\labels$. Here $\curlywedge$ denotes that nothing has been sent or received on $\proc p.c$, $\bullet$ denotes that a value has been sent or received (we do not care about which value), and $l$ denotes that the specific label $l$ has been sent or received. $R$ is a list of recursive procedures and abstract frame dictionaries $\call{X}{\vec{q}}{\vec{k}}:F\rightarrow F'$ where $F$ and $F'$ are the state of the abstract frame dictionaries before and after the execution of the definition of $\call{X}{\vec{q}}{\vec{k}}$. The rules of our robustness system are presented in \cref{fig:robust}.

\begin{figure}[t] \small
\begin{infrules}
\infrule[\rname [R]{Com}][rule:r-com]
	{\rbst{P}{R}{\gnewframepair;C}{F_0}{F_1}}
	{\rbst{P,(\proc p.k,\proc q.k')}{R}{C}{F_0,\proc p.k:\{\curlywedge\},\proc q.k':\{\curlywedge\}}{F_1,\proc p.k:U,\proc q.k':U'}} 
\infrule[\rname [R]{Frame1}][rule:r-frame1]
	{\rbst{P,(\proc p.k,\proc q.k')}{R}{\gnewframe;C}{F_0,\proc p.k:\{\curlywedge\}}{F_1}}
	{\rbst{P,(\proc p.k,\proc q.k')}{R}{C}{F_0,\proc p.k:\{\curlywedge\}}{F_1}} 
\infrule[\rname [R]{Frame2}][rule:r-frame2]
	{\rbst{P,(\proc p.k,\proc q.k')}{R}{\newframe{k'}{q}{p}{T};C}{F_0,\proc q.k':\{\curlywedge\}}{F_1}}
	{\rbst{P,(\proc p.k,\proc q.k')}{R}{C}{F_0,\proc q.k':\{\curlywedge\}}{F_1}} 
\\
\infrule[\rname [R]{Assign}][rule:r-assign]{\rbst{P}{R}{\gassign;C}{F_0}{ F_1}}{\rbst{P}{R}{C}{F_0}{F_1}} \;\;\;\; 
\infrule[\rname [R]{Nil}][rule:r-nil]{\rbst{P}{R}{\cnil}{F}{F}}{} 
\infrule[\rname [R]{Recv}][rule:r-recv]{\rbst{P,(\proc q.c',\proc p.c)}{R}{\grecv;C}{ F_0,\proc p.c:\{\curlywedge\},\proc q.c':U\cup \{u\}}{F_1}}{\rbst{P,(\proc q.c',\proc p.c)}{R}{C}{F_0,\proc p.c:\{\curlywedge,u\},\proc q.c':U }{F_1} \and u\neq \curlywedge \and U\subseteq \{\curlywedge\}}
\infrule[\rname [R]{Send1}][rule:r-send1]{\rbst{P,(\proc p.c,\proc q.c')}{R}{\gsend;C}{ F_0,\proc p.c:\{\curlywedge\}}{F_1}}{\rbst{P,(\proc p.c,\proc q.c')}{R}{C}{F_0,\proc p.c:\{\curlywedge,u\}}{F_1} \and s=l\Rightarrow u=s \and s=e \Rightarrow u=\bullet}
\infrule[\rname [R]{Send2}][rule:r-send2]{\rbst{P,(\proc p.c,\proc q.c')}{R}{ \gsend;C}{F_0,\proc p.c:U}{F_1}}{\rbst{P,(\proc p.c,\proc q.c')}{R}{C}{F_0,\proc p.c:\{\curlywedge,u\}}{F_1} \and s=l\Rightarrow u=s \and s=e \Rightarrow u=\bullet \and U\subseteq \{\curlywedge,u\}}
\infrule[\rname [R]{Weaken}][rule:r-weaken]{\rbst{P}{R}{C}{F_0,F_0'}{F_1,F_1'}}{\rbst{P}{R}{C}{F_0}{F_1}}
\infrule[\rname [R]{Condloc}][rule:r-condloc]{\rbst{P}{R}{\cond{\proc p.e}{C_1}{C_2}}{F_0}{F_1 \mrgf F_2}}{\rbst{P}{R}{C_1}{F_0}{F_1} \and \rbst{P}{R}{C_2}{F_0}{F_2}}
\infrule[\rname [R]{Condsnd}][rule:r-condsnd]{\rbst{P}{R}{\cond{\proc p.c!}{C_1}{C_2}}{F_0,c:\{\curlywedge,u\}}{F_1 \mrgf F_2}}{\rbst{P}{R} {C_1}{F_0,c:\{u\}}{F_1} \and \rbst{P}{R}{C_2}{F_0,\proc p.c:\{\curlywedge\}}{F_2}}
\infrule[\rname [R]{Condrcv}][rule:r-condrcv]{\rbst{P}{R}{\cond{\proc p.c?}{C_1}{C_2}}{F_0,c:\{\curlywedge,u\}}{F_1 \mrgf F_2}}{\rbst{P}{R}{C_1}{F_0,c:\{u\}}{F_1} \and \rbst{P}{R}{C_2}{F_0,\proc p.c:\{\curlywedge\}}{F_2}}
\infrule[\rname [R]{Condlbl}][rule:r-condlbl]{\rbst{P}{R}{\cond{\proc p.c? l}{C_1}{C_2}}{F_0,c:\{\curlywedge,l\}}{F_1 \mrgf F_2}}{\rbst{P}{R}{C_1}{F_0,c:\{l\}}{F_1} \and \rbst{P}{R}{C_2}{F_0,\proc p.c:\{\curlywedge\}}{F_2}}
\infrule[\rname [R]{Cont}][rule:r-cont]{\rbst{P}{R}{\gcont;C}{F_0}{F_1}}{\rbst{P}{R}{C}{F_0}{F_1}} 
\infrule[\rname [R]{Call}][rule:r-call]{\rbst{P}{R,\call{X}{\vec{q}}{\vec{k}}:F\rightarrow F'}{\gcall;C} {F[\vec{\proc p}/\vec{\proc q}][\vec{c}/\vec{k}]}{F''}}{\rbst{P}{R,\call{X}{\vec{q}}{\vec{k}}}{C}{F'[\vec{\proc p}/\vec{\proc q}][\vec{c}/\vec{k}]}{F''}}
\end{infrules}\caption{Robustness rules}\label{fig:robust}
\end{figure}

\Cref{rule:r-com} adds the pair of frames to the list of frames and to the abstract frame dictionary with the only possible payload starting the rest of the choreography being empty. 

In a choreography where one process has already declared their part of a frame, we assume that the pair has been added to $P$ already, and \cref{rule:r-frame1,rule:r-frame2} reflect that. 

\Cref{rule:r-recv} specifies that a receive operation takes any network where the sender may have produced a payload which has not been consumed yet, and yields a network where the payload may be consumed.
In particular, choreographies where receives cannot be matched to sends (e.g., $\ginit;\recv{q}{k'}{x}$) or which have consecutive receive operations (e.g., $\recv{p}{(\proc q,1)}{x};\recv{p}{(\proc q,1)}{y}$) are rejected since delivery is either impossible or inconsistencies may arise (e.g., the second operation shadows a successful outcome of the first).
Given the same receive statement multiple robustness judgements can be derived under the same environment, even once weakening is taken into account. Likewise for send operations (discussed below).
For instance, judgement 
$\rbst{(\proc p.(\proc q,1),\proc q.(\proc p,1))}{\emptyset}{\recv{q}{(\proc p,1)}{x}}{ \proc p.(\proc q,1):{\{\curlywedge,\bullet\}},\proc q.(\proc p,1):{\{\curlywedge\}}}{ \proc p.(\proc q,1):{\{\curlywedge,\bullet\}},\proc q.(\proc p,1):{\{\curlywedge,\bullet\}}} $
is derivable if and only if 
$\rbst{(\proc p.(\proc q,1),\proc q.(\proc p,1))}{\emptyset}{\recv{q}{(\proc p,1)}{x}}{ \proc p.(\proc q,1):{\{\bullet\}},\proc q.(\proc p,1):{\{\curlywedge\}}}{\proc p.(\proc q,1):{\{\bullet\}},\proc q.(\proc p,1):{\{\curlywedge,\bullet\}}} $ is derivable.

\Cref{rule:r-send1,rule:r-send2} are intended for checking robustness in settings without and with omission failures, respectively. The rules both add a possible payload to the sending frame in the form of either $\bullet$ or a specific label depending on payload.
Both are alike save for the requirements imposed on the sender stack: both rules require that the payload is added to the abstract frame dictionary as a possible payload, but only those from the former require that the stack has yet to accept a payload for the frame. 
The more stringent \cref{rule:r-send1} forbids any send operation for a frame with a potentially accepted (hence transmitted) payload (e.g., $\send{p}{k}{e}\conc\send{p}{k}{e'}$) since this is a programming error only if transmission is not guaranteed. In fact, when we have omission failures this has to be programmed at the application level e.g., by resending frames not acknowledged.

\Cref{rule:r-condloc,,rule:r-condsnd,,rule:r-condrcv,,rule:r-condlbl} require branches to be live; graphs are intersected to remove connections created only in one branch; frame dictionaries are merged ($\Ydown$) by pointwise union under the condition that whenever both branches specify a payload they agree on it either being a value or a label i.e., whenever $U_1$ and $U_2$ are merged it must hold that the set $(U_1 \cup U_2)$ is a subset of either $\{\bullet,\curlywedge\}$ or $\{\curlywedge\}\cup\labels$. The interplay between our rules for conditionals and send/receive is key to ensuring at most once communication. \Cref{rule:r-recv,rule:r-send1} both require that the frame has no possibility of a successful communication, and then adds such a possibility to the abstract frame dictionary. The only way to remove this potentially successful communication attempt and be allowed to make another attempt is to enter the else branch of the associated conditional. So ``$\grecv;\grecv;\cnil$" cannot be robust, but ``$\grecv;\cond{\proc p.c?}{\cnil}{\grecv;\cnil}$" can.

\Cref{rule:t-call} simply takes the given definition of the abstract frame dictionaries of the procedure and performs a substitution.

Robustness judgements for procedure definitions are derived using the rule:
\begin{infrules}
	\infrule[\rname[R]{Defs}][rule:r-defs]{
		R;P \vdash \cdefs
	}{
		(\cdefs(X(\vec{\proc q};\vec{k}))=C\wedge R=R',X(\vec{\proc q};\vec{k}):F_1\to F_2)) \Rightarrow P;R \vdash \{F_1\}C\{ F_2\}
	}
\end{infrules}

And robustness judgements for memory configurations are derived using the rule:
\begin{infrules}
	\infrule[\rname[R]{Mem}][rule:r-mem]{
		R;P \vdash \Sigma: F
	}{\begin{array}{c}
	\Sigma(\proc p.\con .\proc q)>i\Rightarrow \{\left(\proc p.\left(\proc q,i\right),\proc q.\left(\proc p,i\right)\right),\left(\proc q.\left(\proc p,i\right),\proc p.\left(\proc q,i\right)\right)\}\cap P\neq \emptyset
		\\ \Sigma(\proc p.\fbuf.\proc q.m) \text{ defined }\Rightarrow \exists F'.F=F',\proc p.(\proc q,m):U \wedge U\in \mathsf{abstract}(\Sigma(\proc p.\fbuf.\proc q.m))
		\\ \Sigma(\proc p.\fbuf.\proc q.m) \text{ defined }\Rightarrow \exists F'. F=F',\proc p.(\proc q,m):U \wedge U\in \mathsf{abstract}(\Sigma(\proc p.\fbuf.\proc q.m))
	\end{array}		
	}
\end{infrules}
This rule uses the function $\mathsf{abstract}$ to generate the acceptable abstract frames corresponding to a frame status. $\mathsf{abstract}$ is defined as: 
\begin{align*}
\mathsf{abstract}(\bot) \defeq{}&\{U\mid \curlywedge \in U\} &
\mathsf{abstract}(\checkmark) \defeq{}& \{U\mid U\cap(\{\bullet\}\cup \labels)\neq \emptyset\} \\
\mathsf{abstract}(v) \defeq{}& \{U\mid \curlywedge \in U\} &
\mathsf{abstract}(l\checkmark) \defeq{}& \{U\mid l\in U\} \\
\mathsf{abstract}(e\checkmark) \defeq{}& \{U\mid \bullet\in U\}
\end{align*}

We can now define robustness of choreographies. The first four items simply require that the configuration can be typed and judged robust. The last item is tricky. It requires that whenever we have a message in transit, the sender and receiver frames have dual input and output types, and those types fit the value of the message. In addition, these frames must in the initial frame directory be respectfully an output frame that has attempted to send the message, and an input frame that has not received anything. 
\begin{definition}[Robustness]
\label{def:rypedness}
Given a runtime configuration, $\langle C, \Sigma, K \rangle$ and a set of procedures, $\cdefs$, we say that $\langle C, \Sigma, K\rangle$ is \emph{robust} under $\cdefs$ if there exist $\Gamma$, $P$, $R$, $F_0$, and $F_1$ such that 
\begin{itemize}
\item $\langle C, \Sigma, K \rangle$ is well-typed using $\Gamma$;
\item $\Gamma\vdash \cdefs$
\item $R;P\vdash \Sigma:F_0$;
\item $R;P \vdash \cdefs$;
\item $\rbst{P}{R}{C}{F_0}{F_1}$;
\item If $\{\left(\proc p.\left(\proc q,i\right),\proc q.\left(\proc p,k\right)\right),\left(\proc q.\left(\proc p,k\right),\proc p.\left(\proc q,i\right)\right)\}\cap P$ and $\{\left(\proc p.\left(\proc q,j\right),\proc q.\left(\proc p,k'\right)\right),\left(\proc q.\left(\proc p,k'\right),\proc p.\left(\proc q,j\right)\right)\}\cap P$ for $i<j$ then in $C$, $\newframe{k}{q}{p}{T}$ appears before $\newframe{k'}{q}{p}{T'}$.
\item if a procedure $X$ is called in $C$ or in a procedure in $\cdefs$, then it is defined in $\cdefs$ with the same number of processes and frames; and
\item for any $\langle \proc p,\proc q,m,v\rangle\in K$ we have $(\proc p.(\proc q,m),\proc q.(\proc p,m))\in P$, a type $T$ such that $\Gamma(\proc p.(\proc q,m))=\sndt{T}$, $\Gamma(\proc q.(\proc p,m))=\rcvt{T}$, $\Gamma(\proc p)\vdash v:T$, and $F_0=F_0',\proc p.(\proc q,m):U\cup\{u\},\proc q.(\proc p,m):\{\curlywedge\}$ where if $v=l$ then $u=l$ and otherwise $u=\bullet$ and $U\subseteq\{\curlywedge\}$.
\end{itemize}

\end{definition}

For any configuration there is an environment and a finite set of pairs of abstract frame dictionaries that subsume any other robustness judgement.

\begin{theorem}[Existence of minimal robustness]
\label{thm:chor-minimal-type}
Let $\langle C, \Sigma, K \rangle$ be \emph{robust} under $\cdefs$. There are $P$, $F_0$, \dots, $F_n$, $F'_0$, \dots, $ F'_n$ with the property that
whenever $\rbst{P}{R}{C}{F}{F'}$ there is $i \leq n$ s.t.:
\[
	\infer[\ref{rule:r-weaken}]{
		\rbst{P}{R}{C}{F_i}{F'_i} 
	}{
		\rbst{P}{R}{C}{F}{F'}
	}
\]
\end{theorem}
\begin{proof}[Proof sketch]
Follows from there existing a finite number of frames mentioned in $C$ and a finite number of possible states of frames in abstract frame dictionaries.
\end{proof}

\begin{theorem}[Decidability of robustness]
\label{thm:decidability}
Given a set of procedure definitions $\cdefs$ and a runtime configuration $\langle C, \Sigma, K \rangle$, it is decidable whether $\langle C, \Sigma, K \rangle$ is robust under $\cdefs$.
\end{theorem}
\begin{proof}[Proof sketch]
Typing for elements of the guest language is known by assumption.
Observe that building derivations for typing judgements from \cref{def:rypedness} is completely mechanical:
rule selection is deterministic save \cref{rule:r-weaken} which only introduces a finite 
number of cases and can be used a finite amount of times. A heuristic is to delay uses of these rules until it is 
necessary for \cref{rule:r-call}. Hence, 
derivations can be built using straightforward non-deterministic case exploration. 
The only nontrivial part is constructing typing environments but only a finite number of cases need 
to be checked since processes, frames, and labels that need to be part of the typing environment 
are inferred from free names in program terms and formal parameters.
Finally, observe that the definition domain of memory configurations and concrete frame 
dictionaries is bounded by typing environments and network types, and that actual values in $\Sigma$ 
or $K$ (the only possible source of infinity) are irrelevant provided that they are of the right 
types (which is checkable by the assumptions on the guest language).
\end{proof}

Robustness is preserved by transitions.
\begin{theorem}[Robustness preservation]
	\label{thm:type-preservation}
	Assume $\langle C,\Sigma, K \rangle$ is robust under $\cdefs$ and there is a transition $\langle C,\Sigma, K \rangle \lto{\mu} \langle C',\Sigma', K' \rangle$. Then the resulting configuration $\langle C',\Sigma', K' \rangle$ is robust under $\cdefs$.
\end{theorem}
\begin{proof}[Proof sketch]
Proved by induction on the transition $\langle C,\Sigma, K \rangle \lto{\mu} \langle C',\Sigma', K' \rangle$. The $\Gamma$, $P$, $F_0$ and $F_1$ used in the robustness judgement get updated as expected.
\end{proof}

Frames are never delivered (to the receiver application level) more than one time. More specifically, if we can send on a certain frame, then that frame's status in the memory does not have a checkmark indicating that a send on the frame has already been executed.
\begin{theorem}[At-most-once delivery]
\label{thm:chor-at-most-once}
Let $\langle C,\Sigma,K \rangle$ be robust under $\cdefs$.
If $\langle C,\Sigma,K \rangle\lto{\rightarrow \proc p.(\proc q.m)}$, then $\Sigma(\proc p.\fbuf.\proc q.m)\neq v\checkmark$.
\end{theorem}
\begin{proof}[Proof sketch]
We prove this by structural induction on $C$. The key case is $C=\gsrecv;C'$. Here the result follows from the definition of $R;P\vdash \Sigma:F$.
\end{proof}

In a setting without omission failures (i.e., \cref{rule:c-loss} cannot be applied to derive transitions), robustness identifies frames that are guaranteed to be delivered. Specifically, if a frame has a chance of a message being delivered on it, then when the choreography reaches a point where the frame will no longer be touched, the message will have been delivered.
\begin{theorem}[At-least-once delivery]
\label{thm:chor-at-least-once}
Assume that there are no omission failures, that the runtime configuration $\langle C,\Sigma, K  \rangle$ is robust under $\cdefs$, and that
$\rbst{P}{R}{C}{F_0}{F_1}$.
For $\proc p.c:U \in F_1$ such that $(\proc q.c',\proc p.c)\in P$ and $\curlywedge \notin U$, 
if $\langle C,\Sigma, K \rangle \lto{\mu_1}\dots \lto{\mu_n} \langle C',\Sigma', K' \rangle$ and $c \notin \mathrm{fn}(C')$ then $\Sigma'(\proc p.\fbuf.c)=v\checkmark$.
\end{theorem}
\begin{proof}[Proof sketch] The only ways $\curlywedge\notin U$ are if $U$ is generated by the left branch of a conditional $\cond{\proc p.c?}{C_0}{C_1}$, which can only be entered if  $\Sigma(\proc p.\fbuf.c)=v\checkmark$.
\end{proof}

There is always an execution where a given frame is delivered.
\begin{theorem}[Best-effort delivery]
\label{thm:chor-best-effort}
Assume possibility of omission failure and the that runtime configuration $\langle C,\Sigma, K  \rangle$ is robust under $\cdefs$, and that
$\rbst{P}{R}{C}{F_0}{F_1}$.
For any $(\proc p.c,\proc q.c')\in P$ such that $\proc p.c:\{\curlywedge\} \in F_0$, $\proc p.c:U \in F_1$, and $U\neq \{\curlywedge\}$, 
if there is a sequence of transitions $\langle C,\Sigma, K \rangle \lto{\mu_1}\dots \lto{\mu_n} \langle C',\Sigma', K' \rangle\ltoc{\rightarrow \proc q.c'}{}$ such that $c \notin \mathrm{fn}(C')$, then 
there exist $C''$, $\Sigma''$, $K''$ such that $\langle 
C,\Sigma,K \rangle \lto{\mu_1}\dots \lto{\mu_n} \langle C'',\Sigma'',K'' \rangle\lto{\rightarrow \proc q.c'}$ and $\Sigma''(\proc q.\fbuf.c')=v$.
\end{theorem}
\begin{proof}[Proof sketch]
Since $U$ has something added to it between $F_0$ and $F_1$, there must be a send on $\proc p.c$ somewhere in $C$, and from \cref{rule:r-recv}, we know that at least one such send must be before the first receive on $\proc q.c'$. The result then follows from the semantics.
\end{proof}

We have shown that robustness guarantees some important properties of communications in choreographies, but note that we only reason about individual communications. Compare \code{sendExpBackoff} from \cref{ex:21} with \code{sendAny} from \cref{sec:examples}. Both guarantee that one successful send operation must occur, so we look at their robustness judgements: $$\rbst{P}{R}{\code{sendExpBackoff}_{e}(\proc p;k)}{\proc p.k:\{\curlywedge\}}{\proc p.k:\{\bullet\}}$$ $$\rbst{P}{R}{\code{sendAny}_{n,e}(\proc p;k_1,\dots,k_n)}{\proc p.k_1:\{\curlywedge\},\dots, \proc p.k_n:\{\curlywedge\}}{\proc p.k_1:\{\curlywedge,\bullet\},\dots, \proc p.k_n:\{\curlywedge,\bullet\}}$$ The former shows that the send must have succeeded, but since each individual frame in \code{sendAny} could have either succeeded or failed, the robustness does not show that a send is guaranteed.

\section{Compiling Processes from Choreographies}
\label{sec:synthesis}

In this section we present an EndPoint Projection (EPP) procedure which compiles a choreography to a concurrent implementation represented in terms of a process calculus. This calculus assumes the same failure model assumed for the choreography model.

\subsection{Lossy Processes}
The target process model is based on Recursive Processes, the textbook target for choreography projection~\cite{M23}. We extend the semantics so send and receive operations may fail and exchanged messages are tagged with numeric identifiers. 
Numeric frame identifiers may be regarded as sequence numbers. However, the model does not offer any mechanism for maintaining counters synchronised among connected processes nor can such mechanism be programmed since these counters are inaccessible. The only way to maintain synchrony is to write programs where frame declarations are carefully matched on each involved party. In a system created from a choreography, the frame declarations are guaranteed to match, as the choreography always declares two corresponding frames at the same time.

\paragraph*{Syntax}
The full syntax of the language for programming in this model is defined by the grammar below.
\begin{align*}
B &\Coloneqq L;B \mid \cnil
	&& L \Coloneqq \ginitl \mid \gsendl \mid \grecvl \mid \gassignl \mid \gcondl \mid \gbranch \mid \gcall\\
	s &\Coloneqq l \mid e \qquad
	&& b \Coloneqq e \mid c \qquad
	c \Coloneqq k \mid (\proc p,m)
\end{align*}
Term $\ginitl $ describes a behaviour that creates a new frame for its continuation. The send action for $c$ is described by term $\gsendl $, and receive by $\grecvl $.
Term $\gbranch $ describes a branching based on a label communicated as frame $k$, if any label $l_i$ has been successfully received then, the process proceeds with the corresponding behaviour $B_i$ otherwise it proceeds with the one labelled with $\lbldefault$. This label is reserved exclusively for this purpose and cannot be sent. If $I = \emptyset$, then the term is simply discarded. Guard $c$ states that the last communication action for frame $c$ has been successfully completed. Remaining terms are standard.

The processes whose behaviour is described by this syntax can be composed into networks:
\begin{definition}
  A network $N$ is a finite map from a set of processes to local programs.
  We often write $\proc{p_1}{B_1} \mid \cdots \mid \proc{p_n}[B_n]$ for the network where process $\proc{p_i}$ has behaviour $B_i$.
\end{definition}

The parallel composition of two networks $\mathcal{N}$ and $\mathcal{N}'$ with disjoint domains,
$N \mid N'$, simply assigns to each process its behaviour in the network defining it.
Any network is equivalent to a parallel composition of networks with singleton domain, as suggested by the syntax above.

\paragraph*{Semantics} 
\begin{figure}[t]\small
\begin{infrules}
	\infrule[\rname[P]{Frame}][rule:p-frame]{
		\tuple{\proc p[\newframel{\proc q}{k} ; B],\Sigma,K}
		\ltol{\proc p\rightarrow \proc q}
		\tuple{\proc p[B[(\proc q,m)/k]],\Sigma',K}
	}{
		m =\Sigma(\proc p.\fcnt.\proc q) \and
		\Sigma' = \Sigma[\proc p.\fcnt.\proc q \mapsto m+1, \proc p.\fbuf.\proc q.m \mapsto \bot]
	} 
	\infrule[\rname[P]{Assign}][rule:p-assign]{
		\tuple{\proc p[\gassignl ; B];\Sigma,K}
		\ltol{\tau@\proc p}
		\tuple{\proc p[B],\Sigma[\proc p.x\mapsto v]}
	}{
		\eval s\Sigma pv
	}
	\infrule[\rname[P]{Snd}][rule:p-send]{
		\tuple{\proc p[\sendl{(\proc q,m)}{s} ; B],\Sigma,K}
		\ltol{\proc p.(\proc q,m) \rightarrow}
		\tuple{\proc p[B],\Sigma',K\cup \{(\proc p,\proc q,m,v)\}}
	}{
		\eval S\sigma pv
		\and
		\Sigma' = \sigma[\proc p.\fbuf.\proc q.m \mapsto \checkmark]
	}
	\infrule[\rname[P]{SndFail}][rule:p-send-fail]{
		\tuple{\proc p[\sendl{(\proc q,m)}{s} ; B],\Sigma,K}
		\ltol{\proc p.(\proc q,m) \rightarrow}
		\tuple{\proc p[B],\Sigma,K}
	}{}
	\infrule[\rname[P]{Recv}][rule:p-recv]{
		\tuple{\proc p[\recvl{(\proc q,m)}{x} ; B],\Sigma,K}
		\ltol{\rightarrow\proc p.(\proc q,m) }
		\tuple{\proc p[B],\Sigma[\proc p.\fbuf.\proc q.m \mapsto v,x\mapsto v\checkmark],K}
	}{
		\Sigma(\proc p.\fbuf.\proc q.m)\in\{v,v\checkmark\}
	}
	\infrule[\rname[P]{RcvFail}][rule:p-rcv-fail]{
		\tuple{\proc p[\recvl{(\proc q,m)}{x} ; B],\Sigma,K}
		\ltol{\rightarrow\proc p.(\proc q,m) }
		\tuple{\proc p[B],\Sigma,K}
	}{ \Sigma(\proc p.\fbuf.\proc q.m)=\bot
	}
	\infrule[\rname[P]{IfFrame}][rule:p-if-frame]{
		\tuple{\proc p[(\cond{(\proc q,m)}{B_1}{B_2});B],\Sigma,K}
		\ltol{\mu }	
		\tuple{\proc p[B_i;B],\Sigma,K}
	}{
		\text{if } \bot\in \{\Sigma(\proc p.\fbuf.\proc q.m), \Sigma(\proc p.\fbuf.\proc q.m)\}
		\text{ then } i = 2 \text{ and } \mu=\mathsf{right}@\proc p
		\text{ else } i = 1 \text{ and } \mu=\mathsf{left}@\proc p
	}
	\infrule[\rname[P]{IfExp}][rule:p-if-exp]{
		\tuple{\proc p[(\cond{e}{B_1}{B_2});B],\Sigma,K}
		\ltol{\mu }	
		\tuple{B_i;B,\Sigma,K}
	}{
		\text{if } \eval s\Sigma p {\literal{true}}
		\text{ then } i = 1 \text{ and } \mu=\mathsf{left}@\proc p
		\text{ else } i = 2 \text{ and } \mu=\mathsf{right}@\proc p
	}
	\infrule[\rname[P]{Branch}][rule:p-branch]{
		\tuple{\proc p[\branch{(\proc q,m)}{l_i:B_i}_{i\in I};B],\Sigma,K}
		\ltol{\mathsf{left}@\proc p}
		\tuple{\proc p[B_i;B],\Sigma,K}
	}{
		\Sigma(\proc p.\fbuf.\proc q.m) = l_i	
		\and
		\lbl{default} \neq l_i
	}
	\infrule[\rname[P]{BranchFail}][rule:p-branch-fail]{
		\tuple{\proc p[\branch{(\proc q,n)}{l_i:B_i}_{i\in I};B],\Sigma,K}
		\ltol{\mathsf{right}@\proc p}	
		\tuple{\proc p[B_i;B],\Sigma,K}
	}{
		\Sigma(\proc p.\fbuf. \proc q.n) = \bot
		\and
		\lbl{default} = l_i
	}
\infrule[\rname[P]{Unfold}][rule:p-unfold]
		{\tuple{\proc p[X\tuple{\procs q;\vec c};B],\Sigma,K} 
		\ltol{\tau@\proc p}
		\tuple{\proc p[B'[\vec{\proc q}/\vec{\proc p}][\vec{c}/\vec{k}];B],\Sigma,K}}
		{\pdefs(X(\vec{\proc p};\vec{k}))=B'}
	\infrule[\rname[N]{Par}][rule:n-par]
		{\tuple{N \apar M,\Sigma,K} \ltol{\mu} \tuple{N'\apar M,\Sigma',K'}}		
		{\tuple{N,\Sigma,K} \ltol{\mu} \tuple{N',\Sigma',K'}}
\infrule[\rname[N]{Loss}][rule:n-loss]{
		\tuple{N,\Sigma,K\cup \{(\proc p,\proc q,m,v)\}} \ltol{\tau} \tuple{N,\Sigma,K}
	}{
	}\infrule[\rname[N]{Com}][rule:n-deliver]{
		\tuple{N,\Sigma,K\cup \{(\proc p,\proc q,m,v)\}} \ltol{\tau} \tuple{N,\Sigma[\proc q.\fbuf.\proc p.m\mapsto v],K}
	}{
	}
\end{infrules}
	\caption{Process model, operational semantics}
	\label{fig:proc-semantics}
\end{figure}
The semantics of networks is given as an LTS.
States are configurations $\tuple{N,\Sigma,K}$ where $N$ is a network and  $\Sigma$ and $K$ are as in the LTS of LC.
The transition relation uses the same labels of LC and is parameterised in a set $\pdefs$ of procedure definitions.

\Cref{rule:p-frame} models the creation at $\proc p$ of a new frame for $\proc q$ using the value $m$ of the corresponding counter.
\Cref{rule:p-send} models the sending of a message $v$ to frame $(\proc q,m)$, updating the memory to mark the frame as sent and \cref{rule:p-send-fail} models the case where the send fails.
Dually, \cref{rule:p-recv,rule:p-rcv-fail} model the successful reception of a message from frame $(\proc q,m)$, updating the memory to mark the frame as received and the variable $x$ with the value $v$, and the case where the receive fails, respectively.
\Cref{rule:n-loss,rule:n-deliver} model the loss and delivery of messages by the network like \cref{rule:c-loss,rule:c-deliver}.

\subsection{EndPoint Projection}
\label{sec:epp}

Given a choreography $C$, the projected behaviour of process $\proc p$ in $C$ is defined as $\epp{C}[\proc p]$ where $\epp{-}[\proc p]$ is the partial function defined by structural recursion in \cref{fig:behaviour-projection}.
Each case in the definition follows the intuition of projecting, for each choreographic term, the local actions performed by the given process.
For instance, $\gsend$ is skipped during the projection of any process but $\proc p$ for which case the send action $\gsendl$ is produced. Cases for receiving, procedure calls, and frame creation are similar. The projection of frame declaration expands it to two frame creation runtime terms similarly to its semantics, ensuring that as long as the starting configuration is consistent, the frame counters will remain synchronised for the entire execution. 
The case for conditionals is more involved but follows a standard approach (see e.g., \cite{BCDLDY08,LGMZ08,CHY12,CM16:facs,CM17:forte}). The (partial) merging operator $\merge$ from \cite{CHY12} is used to merge the behaviour of a process that does not know (yet) which branch has been chosen by the the process evaluating the guard. Intuitively, $B \merge B'$ is isomorphic to $B$ and $B'$ up to branching, where branches of $B$ or $B'$ with distinct labels are also included. One proceeds homomorphically (e.g., $\sendl{k}{e}; B \merge \sendl{k}{e}; B'$ is $\sendl{k}{e}; (B \merge B')$) on all terms but branches which are handled defining the merge of
$
	\branch{c}{l_i : B_i}_{i \in I}; B 
$ and $
	\branch{c}{l_j: B'_j}_{j \in J}; B'
$
as
$\branch{c}{l_h : B''_h}_{h \in H}; (B \merge B')$
where $\{l_h: B''_h\}_{h \in H}$ is the union of
$\{l_i : B_i\}_{i \in I\setminus J}$, $\{l_j : B'_j\}_{j \in J\setminus I}$, and $\{l_g : B_g \merge B'_g\}_{g \in I\cap J}$.

Projection of procedure definitions follows the approach introduced by Procedural Choreographies \cite{CM17:forte}. For $\cdefs$ a set of procedure definitions, its projection is defined as follows:
\[
	\epp{\cdefs} \defeq \bigcup_{X(\proc {p_1},\dots \proc {p_n};\vec{k})=C \in \cdefs}
			\left\{
				X_{i}(\proc {p_1},\dots,\proc p_{i-1},\proc p_{i+1},\dots \proc {p_n};\vec{k}){\epp{C}[\proc p_i]}
			\,\middle|\,
				1\leq i \leq n 
			\right\}
	\text{.}
\]
Observe that since a procedure $X$ may be called multiple times on any combination of its arguments (hence assigning a process to different roles at each call) it is necessary to project the behaviour of each possible process parameter $\proc p_i$ as the procedure $X_{i}$.

\begin{figure}[t]
	\begin{center}
	\begin{autoflow}	
		\newcommand{\afdbox}[1]{{$\displaystyle#1$}\AND}

		\afdbox{
			\epp{\cnil}[\proc r] 
			\defeq \cnil
		}
		
		\afdbox{
			\epp{\gcall;C}[\proc r] 
			\defeq 
			\begin{cases}
				X_{i}(\vec{p}\setminus \proc r;\vec{c});\epp{C}[\proc r] & \text{if } 
				\proc p_i\tteq \proc r \in \vec{\proc p} \text{ for some } i\\
				\epp{C}[\proc r] & \text{otherwise}
			\end{cases}
		}
		
		\afdbox{
			\epp{\gcont;C}[\proc r] 
			\defeq 
			\begin{cases}
				X_{i}(\vec{p}\setminus \proc r;\vec{c});\epp{C'}[\proc r] & \text{if } 
				\proc r \in \procs q \text{ and } \proc p_i\tteq \proc r \in \vec{\proc p} \text{ for some } i\\
				\epp{C}[\proc r] & \text{otherwise}
			\end{cases}
		}
		
		\afdbox{
		\epp{\gnewframepair ;C}[\proc r] 
		\defeq  
		\epp{\gnewframe;\newframe{k'}{q}{p}{T};C}[\proc r]}
		
		\afdbox{
		\epp{\gnewframe;C}[\proc r] 
		\begin{cases}
			\newframel{\proc q}{k};\epp{C}[\proc r] & \text{if } \proc r = \proc p\\
			\epp{C}[\proc r] & \text{else}
		\end{cases}}
		
		\afdbox{
			\epp{\gsend ; C}[\proc r] 
			\defeq 
			\begin{cases}
				\gsendl; \epp{C}[\proc r] & \text{if } \proc r = \proc p\\
				\epp{C}[\proc r] & \text{otherwise}
			\end{cases}
		}
		
		\afdbox{
			\epp{\grecv ; C}[\proc r] 
			\defeq 
			\begin{cases}
				\grecvl ; \epp{C}[\proc r] & \text{if } \proc r = \proc p\\
				\epp{C}[\proc r] & \text{otherwise}
			\end{cases}
		}
		
		\afdbox{
			\epp{\gassign ; C}[\proc r] 
			\defeq 
			\begin{cases}
				\gassignl ; \epp{C}[\proc r] & \text{if } \proc r = \proc p\\
				\epp{C}[\proc r] & \text{otherwise}
			\end{cases}
		}
				
		\afdbox{
			\epp{(\gcond);C}[\proc r] 
			\defeq 
			\begin{cases}
				(\branch{c}{l: \epp{C_1}[\proc r],\lbldefault: \epp{C_2}[\proc r]});\epp{C}[\proc r] & 
					\text{if } \proc p.b=\proc r . c? l\\
				(\cond{e}{\epp{C_1}[\proc r]}{\epp{C_2}[\proc r]});\epp{C}[\proc r] & 
					\text{if } \proc p.b = \proc r.e \\
				(\cond{{c}}{\epp{C_1}[\proc r]}{\epp{C_2}[\proc r]});\epp{C}[\proc r] & 
					\text{if } \proc p.b = \proc r.c! \text{ or } \proc p.b = \proc r.c?\\
				(\epp{C_1}[\proc r] \merge \epp{C_2}[\proc r]); & 	
					\text{otherwise}
			\end{cases}
		}

	\end{autoflow}
	\end{center}
	\caption{Process projection.}
	\label{fig:behaviour-projection}
\end{figure}

Below we show the projection of the implementation of two-phase commit from \cref{sec:2pc} for a participant $\proc {p_i}$:
\begin{snippet}
	\procdef{\code{2PhaseCommit}}{
		\proc c,
		\proc {p_1},
		\dots, \proc {p_{i-1}},\proc {p_{i+1}},\dots
		\proc {p_n};
	\emptyset}{
	\\\indent
		\initl{\proc c}{k_{v_i}};
		\initl{\proc c}{k'_{d_i}}; 
		\initl{\proc c}{k_{a_i}}; \commentline{Create frames for communicating with $c$}
		\\\indent
		\sendl{k_{v_i}}{\mathsf{vote}()}; \commentline{Send vote}
		\\\indent
		\code{RecvDec}(\emptyset; k'_{d_i},k_{a_i}) \commentline{Call procedure for receiving decision}
	}
	\\
	\procdef{\code{RecvDec}}{\emptyset ;k_{d},k_{a}}
	{
	\\\indent
	\recvl{k_{d}}{dec}; \commentline{Try Receiving the decision}
	\\\indent
	\cond{k_{d}}{ \commentline{If receive was succesful}
	\\\indent\indent
	\cond{dec}{ \commentline{execute decision to abort or commit}
	\\\indent\indent\indent
	\assignl{memory}{\mathsf{commit}()}
  \\\indent\indent
	}
	{\\\indent\indent\indent
	\assignl{memory}{\mathsf{abort}()}
	}
	\\\indent\indent
	\code{sendExpBackoff}_{unit}(\emptyset ,k_{a}) 
  \commentline{Keep sending acknowledgements until successful}
  \\\indent
	}
	{ 
	\\\indent\indent
	\code{RecvDec}(\emptyset ;k_{d},k_{a}) \commentline{If receive decision was unsuccessful, try again}
	}
	}
\end{snippet}
Here we see how the projection ignores actions at other processes, but keeps the behaviour $\proc {p_i}$ the same. \code{RecvDec} therefore winds up looking more or less the same as in the choreography, because it takes place entirely at one process, while in \code{2PhaseCommit} we remove the majority of the code, as it takes place at $\proc c$ and the other participants. Note also that $\proc {p_i}$ does not know the names of the frames at $\proc c$ which it is sending to, but thanks to our projection and a consistent memory configuration, $\Sigma$, we can guarantee that the corresponding frames will be instantiated with the same identifier.

There is an operational correspondence between choreographies and their projections, which we can formulate in the standard way
up to the `branching' relation $\pruning$ (\cite{M23}).
This relation accounts for the fact that, after a conditional is executed at the choreography level, some processes get to know about the chosen branch only after a while (through selections) and thus have temporary `dead code' (branches that are never going to be selected).
Formally, the branching relation is defined as $P \pruning Q$ iff $P \merge Q = P$.

\begin{theorem}[EPP]
	\label{thm:epp}
	Given a well-typed configuration $\tuple{C,\Sigma,K}$ and set of procedure definitions $\cdefs$ such that $\epp{C}$ and $\epp{\cdefs}$ are defined, then:
	\begin{itemize}
		\item
			If $\tuple{C,\Sigma,K} \lto{\mu} \tuple{C',\Sigma',K'}$ 
			then
			$\epp{C,\Sigma,K} \ltoc{\mu}{\epp{\cdefs}} N$ 
			such that $\epp{C'} \pruning N$.
		\item
			If $\tuple{\epp{C},\Sigma,K} \ltoc{\mu}{\epp{\cdefs}} \tuple{N,\Sigma',K'}$, 
			then $\tuple{C,\Sigma,K} \lto{\mu} \tuple{C',\Sigma',K'}$ 
			such that $\epp{C'} \pruning N$.
	\end{itemize}
\end{theorem}
\begin{proof}[Proof sketch]
We prove completeness by induction on $\langle C,\Sigma,K \rangle \lto{\mu} \langle C',\Sigma',K' \rangle$ and soundness by induction on $\tuple{\epp{C},\Sigma,K} \ltoc{\mu}{\epp{\cdefs}} N$. Both of these proofs are fairly standard for choreographies with the exception of splitting the send and receive into separate actions and adding the failure transitions, which work the same in the choreography and local semantics.
\end{proof}

It follows from the operational correspondence in \cref{thm:epp} that projected networks exhibit all relevant properties ensured by the robustness introduced in \cref{sec:robust}, namely: progress (\cref{thm:progress}), at-most-once delivery (\cref{thm:chor-at-least-once}), best-effort delivery (\cref{thm:chor-best-effort}), and at-lest-once delivery (\cref{thm:chor-at-least-once}).

\section{Implementing Frames in Choral}
\label{sec:implementation}

\subsection{A taste of Choral}
Choral \cite{GMP24} is an object-oriented choreographic programming language: in Choral classes and interfaces represent distributed data types parametric in the processes (\emph{roles} in Choral terminology) participating in them. 
The syntax of Choral is based on that of Java, the only main difference is the introduction of notation for process parameters (recognisable by \choral{@}) in types as exemplified in the snippet below which contains the definition of a distributed pair storing two values at different roles. 
\begin{chorallisting}[breakable=false][numbers=none]
public class DPair@(#A#,#B#)<L@#C#,R@#D#> {
  public final L@#A# left;  public final R@#B# right;  
  public DPair(L@#A# left, R@#B# right) {
    this.left = left;  this.right = right;
  } }
\end{chorallisting}
The Choral class \choral{DPair} is distributed over two processes represented by the parameters \choral{#A#} and \choral{#B#}.
The definition is also parametrised on two datatypes \choral{L} and \choral{R} which are expected to be parametrised on exactly one process abstracted by \choral{#C#} for \choral{L} and \choral{#D#} for \choral{R}, respectively. 
These type parameters are instantiated in the definitions of the fields  \choral{left} and \choral{right} to locate these at \choral{#A#} (\choral{L@#A#}) and \choral{#B#}.

The Choral compiler generates a Java implementation for each process participating in a Choral type. 
The class \choral{DPair} above is compiled to the following two Java classes.
\begin{center}
\begin{minipage}{0.49\linewidth}
\begin{javalisting}[breakable=false][numbers=none]
// Implementation of DPair for A
public class DPair_A<L,R> {
  public final L left;
  public DPair_A(L left) {
    this.left = left;
  } }
\end{javalisting}
\end{minipage}\hfill\begin{minipage}{0.49\linewidth}
\begin{javalisting}[breakable=false][numbers=none]
// Implementation of DPair for B
public class DPair_B<L,R> {
  public final R right;  
  public DPair_B(R right) {
    this.right = right;
  } }
\end{javalisting}
\end{minipage}
\end{center}

Differently from other choreographic models and programming languages, Choral does not fix a communication primitive nor a middleware. Instead, communication mechanisms can be programmed directly.
The choral standard library provides a framework with several kinds of channels organised around a hierarchy of interfaces that document standard capabilities (e.g., uni- or bidirectional channels). 
Programmers can thus rely on standard implementations provided by Choral or deploy their own solutions written directly in Choral or in Java (\cite[Sec.~2.4]{GMP24}). 
The snippet below contains the interface defined by the Choral standard library to represent a generic directed channel between two processes (abstracted by \choral{#A#} and \choral{#B#}) for transmitting data of a given type (abstracted by the type parameter \choral{#T#}) from the first to the second --- as for \choral{DPair}, \choral{#C#} and \choral{#D#} are used to abstract the role in the declaration of the type variables \choral{T} and \choral{M}. 
\begin{chorallisting}[breakable=false][numbers=none]
public interface DiDataChannel@(#A#,#B#)<T@#C#> {
  <M@#D# extends T@#D#> M@#B# com(M@#A# message); 
}
\end{chorallisting}
Data transmission is performed by invoking the generic method \choral{com} which takes any value of a subtype \choral{M} of \choral{T} located at \choral{#A#} and returns a value of the same type but located at \choral{#B#}. 
The interface \choral{DiDataChannel} is compiled to the following two Java interfaces, one for each process.
\begin{center}
\begin{minipage}{0.49\linewidth}
\begin{javalisting}[breakable=false][numbers=none]
// Implementation of DiDataChannel for A
public interface DiDataChannel_A<T> {
  <M extends T> void com(M message); 
}
\end{javalisting}
\end{minipage}\hfill\begin{minipage}{0.49\linewidth}
\begin{javalisting}[breakable=false][numbers=none]
// Implementation of DiDataChannel for B
public interface DiDataChannel_B<T> {
  <M extends T> M com(); 
}
\end{javalisting}
\end{minipage}
\end{center}

The classical primitive $\proc p.5 \to \proc q.x$ found in many models of choreographic programming can be realised in Choral as
the statement below---we assume that the variable $\proc q.x$ is fresh and thus requires a declaration which in turn provides us with a chance to illustrate how data location is tracked with types.
\begin{chorallisting}[breakable=false][numbers=none]
Integer@#Q# x = ch.<Integer>com(5@#P#);
\end{chorallisting}
This statement is compiled to the following two Java statements, one to be executed by \choral{#P#} and one to be executed by \choral{#Q#}.
\begin{center}
\begin{minipage}{0.49\linewidth}
\begin{javalisting}[breakable=false][numbers=none]
// Implementation P
ch.<Integer>com(5@);
\end{javalisting}
\end{minipage}\hfill\begin{minipage}{0.49\linewidth}
\begin{javalisting}[breakable=false][numbers=none]
// Implementation for Q
Integer x = ch.<Integer>com();
\end{javalisting}
\end{minipage}
\end{center}

More examples can be found in the original Choral paper \cite{GMP24} and in \cite{PPM24} which introduced a library for non-blocking communication in Choral.
However, all published libraries assume a reliable setting. 

To support the definition of new communication mechanisms, the Choral standard library also contains a host of utilities for handling common tasks such as serialisation to various text and binary formats that we will utilise in ours implementation of frames. 

\subsection{Frames in Choral}
We realise the primitive for creating a frame ($\init{k}{k'}{\proc p}{\proc q}{\typef{T}}$) as a factory and a pair of frames, one for the sender and one for the receiver.
Following the design of the Choral channel framework, we represent this pattern by defining the generic interfaces below. 
\begin{center}
\begin{minipage}{0.49\linewidth}
\begin{chorallisting}[breakable=false][numbers=none]
public interface FrameFactory@(#A#,#B#)<T@#C#> {
  <M@#C# extends T@#C#> FramePair@(#A#,#B#)<M> com();
}

public class FramePair@(#A#,#B#)<M@#C#> {
  public final SenderFrame@#A#<M> sender;
  public final ReceiverFrame@#B#<M> receiver;
  /* ... */
}
\end{chorallisting}
\end{minipage}\hfill\begin{minipage}{0.49\linewidth}
\begin{chorallisting}[breakable=false][numbers=none]
public interface SenderFrame@#A#<M@#C#> {
  void send(M@#A# m);
  boolean@#A# sent();
}

public interface ReceiverFrame@#A#<M@#C#> {
  M@#A# receive();
  boolean@#A# received();
}
\end{chorallisting}
\end{minipage}
\end{center}
The design of these types purposely follows precisely that of LC and recovers the guarantees offered by its type system
via Choral type system. Using a factory ensures that sender and receiver frames are created in pairs which is crucial
to achieve fully distributed implementations that rely only on local state to maintain the correlation of messages
and frames, as discussed for LC. 
To support Java modern asynchronous programming style, \choral{ReceiverFrame} could additionally extend the interface \java{CompletionStage} from \java{java.util.concurrent} (we omit this for brevity).

The classes that realise these interfaces rely on Java NIO byte channels (hence the prefix \choral{Byte} in their names), specifically on \java{DatagramChannel} for UDP transport and the suite of binary serialisers provided by Choral.
For convenience of exposition, we assume that the size of payloads is within the datagram size and forego multipacket transmission. 

The interface \choral{FrameFactory} is realised by the \choral{ByteFrameFactory} found in the snippet below.
\begin{chorallisting}[breakable=true][]
public class ByteFrameFactory@(#A#,#B#)<T@#C#> implements FrameFactory@(#A#,#B#)<T>{
  // Communication support
  private final ChoralSerializer@A<T,ByteBuffer> serializerA;
  private final ChoralSerializer@B<T,ByteBuffer> serializerB;
  private final WritableByteChannel@#A# channelA;
  private final Dispatcher@B<T> dispatcherB;
  // counters for generating frame identifiers at A and B
  private int@#A# cntA = 0;
  private int@#B# cntB = 0;
  
  public ByteFrameFactory(
      ChoralSerializer@A<T,ByteBuffer> serializerA, ChoralSerializer@B<T,ByteBuffer> serializerB,
      WritableByteChannel@#A# channelA, ReadableByteChannel@#B# channelB 
    ) {
      this.serializerA = serializerA;
      this.serializerB = serializerB;
      this.channelA = channelA;
      this.dispatcherB = new Dispatcher@B(channelB);
      new Thread( dispatcherB ).start();
  }  
  public <M@#C# extends T@#C#> FramePair@(#A#,#B#)<M> newFrame() {
    // generate a pair of frames with fresh identifiers at the sender and receiver
    SenderFrame<M>@#A# sf = new ByteSenderFrame<M>(cntA++, serializerA, channelA);
    ByteReceiverFrame<M>@#B# rf = new ByteReceiverFrame<M>(cntB++, serializerB);
    // instruct the dispatcher to monitor messages for this frame
    dispatcherB.register(rf);
    return new FramePair@(#A#,#B#)<M>(sf,rf);
  }  
  public void close() {
    channelA.close();
    dispatcherB.close();
  } 
}
\end{chorallisting}
Each instance of the factory maintains a pair of binary channels for transmitting data from the sender to the receiver and a thread for listening for packets at the receiver side and dispatch their payload to the object representing the corresponding frame at the receiver end. 
The correlation between payloads and frames is represented using integer identifiers that assigned by the factory when a pair of frames is created.
These identifiers are generated during the invocation of the factory method \choral{newFame} by reading and incrementing the counters \choral{cntA} and \choral{cntB} located at \choral{#A#} and \choral{#B#}, respectively without any coordination so it is crucial that these counters are kept aligned.
By implementing this factory in Choral we can leverage its compilation to ensure this requirement is met: when the factory is used by choreographic code the counters are initialised to the same value when the constructor is invoked and are incremented consistently when the factory method \choral{newFame} is invoked. 
This solution is designed following LC and its EPP. 

The classes \choral{Dispatcher}, \choral{ByteSenderFrame}, and \choral{ByteReceiverFrame} are located at a single process and are implemented directly in Java (this allows us to take advantage of the wildcard \java{?} in generics since this feature is currently not available in Choral).
Below, we report the the main parts and omit constructors and accessors when these are clear from the context. 
\begin{javalisting}[breakable=true][]
import choral.runtime.serializers.ChoralSerializer;
import java.nio.channels.ReadableByteChannel;
import java.nio.channels.WritableByteChannel;
import java.nio.ByteBuffer;
import java.util.Map;
import java.util.concurrent.ConcurrentHashMap;

class Dispatcher<T> implements Runnable { 
  private final ReadableByteChannel channel;
  private final Map<Integer, ByteReceiverFrame<? extends T>> frames = new ConcurrentHashMap<>();
  /* ... */
  public void run() {
    while(channel.isOpen()) {
      ByteBuffer buffer = ByteBuffer.allocate(MTU);
      channel.read(buffer);
      int id = buffer.getInt();
      ByteReceiverFrame<?> r = pending.remove(id);
      if (r != null) r.process(buffer);
    } 
  }  
  void register(ByteReceiverFrame<? extends T> r) {
    pending.put(r.id(),r);
  } 
}

class ByteSenderFrame<T> implements SenderFrame<T> {
  private final int id;
  private final ChoralSerializer<? super T, ByteBuffer> serializer;
  private final WritableByteChannel channel;
  private boolean sent = false;
  /* ... */
  public void send(T m) {
    ByteBuffer b = serializer.<T>to(new Message<T>(id, m));
    this.sent = channel.write(b) > 0;
  } }

class ByteReceiverFrame<T> implements ReceiverFrame<T> {
  private final int id;
  private final ChoralSerializer<? super T, ByteBuffer> serializer;
  private T payload = null;
  private boolean received = false;
  /* ... */ 
  void process(ByteBuffer buffer) {
    Message<T> m = serializer.<T>from(buffer);
    this.payload = m.payload;
    this.received = true;
  }
  public T receive() {
    return this.payload;
  }
  public boolean received() {
    return this.received;
  } 
}

class Message<M> {
  final int id;
  final M payload;
  /* ... */
}
\end{javalisting}

The Choral interface \choral{FrameFactory} fixes the direction of the communication from \choral{#A#} to \choral{#B#} and the snippet below extends it to support communications in the opposite direction. 
\begin{chorallisting}[breakable=true][]
public interface SymFrameFactory@(#A#,#B#)<T@#C#> extends FrameFactory@(#A#,#B#)<T> {
  FrameFactory@(#B#,#A#)<T> flip();
  default <M@#C# extends T@#C#> DPair@(#B#,#A#)<SenderFrame<M>,ReceiverFrame<M>> newSymFrame() {
    return flip().<M>newFrame();
  } }
\end{chorallisting}
Ideally, one would want to mimic the design of symmetric channels in the Choral standard library with are defined as a subtype of directed channels in both directions.
\begin{chorallisting}[breakable=true][]
interface SymDataChannel@(#A#,#B#)<T@#C#> extends DiDataChannel@(#A#,#B#)<T>, DiDataChannel@(#B#,#A#)<T> {
  SymDataChannel@(#B#,#A#)<T> flip();
}
\end{chorallisting}
Applying this approach to \choral{FrameFactory} fails because the two versions of method \choral{newFrame} inherited from \choral{FrameFactory@(#A#,#B#)} and \choral{FrameFactory@(#B#,#A#)} have the same signature but different return types resulting in a conflict.

\subsection{Two Phase Commit} 
To briefly validate the design of our API for Choral we used it to translate our LC implementation of 2PC (\cref{sec:2pc}) into Choral. The translation is essentially 1-1.

For each of the participants \choral{#P1#}, \dots, \choral{#Pn#} the choreography takes as argument an object representing the local transaction at that participant (with methods for committing, aborting, and deciding to vote to commit or abort) and a symmetric frame factory between that participant and the coordinator \choral{#C#}.

\begin{chorallisting}[breakable=true][]
void twoPhaseCommit(SymFrameFactory@(#P1#,#C#) f1, LocalTransaction@#P1# lt1, ...) {
  FramePair@(#P1#,#C#) kv1 = f1.<Boolean>newFrame(); ...          // Frames for casting votes
  FramePair@(#C#,#P1#) kd1 = f1.flip().<Boolean>newFrame(); ... ; // Frames for broadcasting the decision
  FramePair@(#P1#,#C#) ka1 = f1.<Unit>newFrame(); ... ;           // Frames for acknowledgments
  kv1.sender.send( lt1.canCommit() ); ... // Participants send votes
  // Controller receives votes and defaults to 'false' on failure
  Boolean@#C# vote1 = kv1.receiver.received() && kv1.receiver.receive(); ... 
  Boolean@#C# decision = vote1 && ... ;
  // controller broadcasts decision
  sendAllUntilAck(decision, List.of(kd1.sender, ...), List.of(ka1.sender, ...));
  // participants receive the decision and commit/abort their local transaction
  recvDec(kd1.receiver, ka1.sender, lt1), ...
}

void sendAllUntilAck(Boolean@#C# decision, List@#C#<SenderFrame<Boolean>> kdq, List@#C#<ReceiverFrame<Unit>> kaq) {
  if ( kdq.isEmpty() ) return;
  SenderFrame@#C#<Boolean> kd = kdq.remove(0); 
  ReceiverFrame@#C#<Unit> ka = kaq.remove(0);
  kd.send(decision);
  if ( !ka.received() ) {  kdq.add(kd); kaq.add(ka); }
  sendAllUntilAck(decision, kdq, kaq); 
}

void recvDec(ReceiverFrame@#P#<Boolean> kd, SenderFrame@#P#<Unit> ka, LocalTransaction@#P# lt) {
  if ( kd.received() ) {
    if ( kd.receive() ) lt.commit();
    else lt.abort();
    sendExpBackoff(ka, Unit.UNIT );
  } else {
    recvDec(kd,ka,lt);
  } }
\end{chorallisting}

\section{Related Work}
\label{sec:related}

\looseness=-1
The work nearest to ours is \cite{APN17}, which extends multiparty session types \cite{HYC16} -- choreographies used to specify communication protocols abstracting from data and computation.
Specifically, \cite{APN17} adds unreliability to multiparty session types by allowing parts of a choreography declared in special optional blocks to become no-op non-deterministically. 
A static analysis guarantees that the network cannot get  stuck even if all optional blocks are not executed. 
We see our work as complementary: while our initial motivation is similar, the aim is different. 
Our focus is providing guarantees on implementations, and consequently choreographies in LC are concrete programs, in contrast with protocol specifications. 
Another key difference is that communication in \cite{APN17} is synchronous: if a participant succeeds in sending a message, it knows that the receiver has also succeeded. 
In LC, we are interested in systems with asynchronous message passing. 
This requires defining and analysing send and receive actions separately, since succeeding in the former does not necessarily mean succeeding in the latter.
Separating between send and receive actions is also essential to the programming of recovery 
strategies in LC, which may be asymmetric for sender and receiver. For example, a sender may 
have different conditions to check (e.g., a number of retries) than those at the intended receiver 
for deciding whether an action should be retried, which cannot be captured in \cite{APN17}. 
Recovery strategies cannot be specified at all in the choreographies of \cite{APN17}, which is 
another key distinction with our work.
The modelling of recovery strategies in LC is also what allows us to develop our type system for 
the static verification of at-most-once and exactly-once delivery, which is not studied in 
\cite{APN17}.

Another approach to multiparty session types with failures is presented by \cite{VHEZ21}, which instead of communication failures focuses on process failures. Like LC and unlike \cite{APN17}, this framework deals with asynchronous communication, but it still presents some relevant limitations compared to LC. Crucially, \cite{VHEZ21} assumes that certain processes are robust and therefore cannot fail, and that communication between non-failed processes is reliable.
This is in contrast to LC where communication between any processes can reach or not reach the receiver in any order. 
In fact, the failure mode of \cite{VHEZ21} sits between the crash \cite{LF82} and fail-stop \cite{SS83,S84} modes, which impose qualitatively stronger reliability assumptions compared to the omission modes that we consider in LC \cite{PT86,C91}.
Process failures can be encoded in LC, e.g., by having a process making an internal nondeterministic choice to proceed or shut down. Our robustness analysis would then point out that the frames that depend on this choice would not be guaranteed to be delivered.

In \cite{CGY16}, multiparty session types are augmented with controlled exceptions. These are different from communication failures, because they are controlled by the 
programmer and their propagation is ensured through communications that are assumed reliable. 
This approach has been refined in \cite{CVBZE16}, by allowing for more fine-grained propagation of errors (but errors are still user-defined, so similar comments apply). 
A similar approach to session types with exception handling appears in \cite{FLMD19}, though using binary session types on a functional calculus with dynamic topology.

A different approach to multiparty session types with failures is seen in \cite{LNY22}. Like our work, they allow communication to fail nondeterministically, but when this happens they kill the session where the failure occurred. Like \cite{CGY16}, they method of killing the sessions assumes a reliable way to communicate failure between processes.

In \cite{LNN16}, the authors present a choreographic programming model that considers potential failures of 
nodes (processes in our terminology). This approach is far from ours and that in \cite{APN17}, 
since the idea is that a system has redundant copies of a node type, and that a choreography can 
specify how many nodes of a type are needed to continue operating. No recovery can be programmed, 
and there is no presentation of how the approach can be adopted in realistic process models 
(compilation). Communications among functioning nodes are assumed infallible.

Previous work explored a notion of bisimulation for a process calculus with explicit locations and links, where both nodes and links may fail at runtime \cite{FH08}. Differently from our setting, communications are synchronous, messages cannot be lost, and failures are permanent.
Exploring a similar notion of behavioural equivalence for LC and our target process calculus is  definitely an interesting future work, because it may lead to a substitutability principle for generated processes wrt choreographies.
For example, we could replace a block of process code projected from a choreography with an equivalent one without having to re-run compilation.

Previous works on choreographies include choice operators that behave non-de\-term\-in\-ist\-ic\-ally, e.g.,
$C + C'$, read ``run either $C$ or $C'$'' 
\cite{QZCY07,LGMZ08,CHY12,M23} (and their labelled variant, in 
\cite{HYC16}).
These operators do not capture the communication failures that we are interested in, for two 
reasons. First, they are programmed explicitly and are thus predictable.
Second, their formalisations assume that the propagation of choice 
information among processes is reliable (for compilation).
Thus, similar comments to those for the comparison with \cite{CGY16} apply.

We use a syntax reminiscent of Hoare logic \cite{AO19,H69} when reasoning about the communications in our choreography. Previous work used Hoare-like logics to reason about choreographies \cite{CGMP23} and ensure agreement on distributed choices \cite{JV22}. Our work differs in that we are not concerned with the values of variables but the possible success and failure of communications.

Our approach of recovering high-level choreographic constructs by wrapping lower-level communication actions (as in \cref{ex:21}) follows the practice established by Choral, the most expressive implementation of choreographic programming to date~\cite{GMP24}.
Robustness is the first analysis that bridges these two levels of abstraction.
In general, we think that allowing programmers to work with a simple high-level semantics under custom assumptions (e.g., guaranteed quality of service) is an interesting direction for future work.

Recent work explored how choreographic programming languages can be offered as embedded domain-specific languages in, for example, Haskell and Rust~\cite{SKK23,KSZK23,LCH24,SK24}.
Differently from Choral, these languages are based on interpretation rather than compilation.
We think that our ideas are applicable also to these languages, but it would require updating both the choreographic languages and their interpretation functions -- whereas our Choral implementation does not require any change to Choral itself, it is `just a library'.

Our new primitive for communication declaration is reminiscent of the `cut' operator found in session-typed process calculi for connecting two endpoints of a channel~\cite{V12,CLMSW16,KMP19}.
There is a superficial similarity in that, like endpoints, our frames are always created in pairs.
However, endpoint pairs represent channels whereas frame pairs represent single communications.
More importantly, the dynamic creation of endpoints requires synchronisation, whereas creating our new frames does not.
This is essential to our development, since communication is unreliable.
Another difference is that session types typically guarantee (system-wide) progress by restricting the structures of communications or the connection topology of processes. Instead, we extend the standard property of choreographic programming to the setting of unreliable communication: progress does not require such restrictions.

\section{Conclusion}
\label{sec:concl}

Choreographic programming has been investigated for more than a decade, but always under the strong assumption of reliable communication~\cite{M13:phd,CM13,Hetal16,DGGLM17,HG22,JV22,M23,SKK23,KSZK23,GMP24}.
Our study frees the paradigm from this assumption, reaching all the way to standard failure modes and realistic distributed protocols.

Our work covers all failure modes with honest participants.
A natural next step is therefore to investigate how to take into account malicious participants and the general setting of Byzantine failures.
Exploring this direction would touch on the realm of agreement protocols, where possible guarantees are more limited. Therefore, it would be interesting and challenging to understand how close choreographic programming can be brought to the theoretical limit.

Another interesting direction is to explore a quantitative semantics and quantitative properties of Lossy Choreographies.
For instance, in a probabilistic settings, failures are characterised by probability distributions and properties like progress, at-most-once, and exactly-once delivery are formalised as almost-certain properties (their complement event has null measure).
This would enable reasoning about aspects such as quality of service, throughput, probability of critical failure, expected number of retransmissions, etc.

Our static analyses can guarantee that the distributed implementations generated by endpoint projection (in terms of processes) have at-most-once or exactly-once delivery guarantees.
These guarantees are limited to single communications but our approach can be reasonably extended to communication groups. 
This extension has immediate applications, e.g., to the statical verification of replication protocols where an update is deemed successful only if it accepted by enough replicas.

\bibliography{biblio}

\end{document}